\documentclass[12pt]{article}
\usepackage[english]{babel}
\usepackage[iso-8859-7]{inputenc}
\usepackage{verbatim}
\usepackage{amsmath,amsfonts,amssymb,amsthm,url}
\usepackage{epsfig}
\usepackage[blocks,auth-lg,affil-it]{authblk}
\usepackage[hang,nooneline]{subfigure}
\newcounter{fig}   

\usepackage{upref}

\newcommand{\be}{\begin{equation}}
\newcommand{\ee}{\end{equation}}
\newcommand{\bea}{\begin{eqnarray}}
\newcommand{\eea}{\end{eqnarray}}
\newcommand{\bean}{\begin{eqnarray*}}
\newcommand{\eean}{\end{eqnarray*}}

\newcommand{\fr}{\frac}

\newcommand{\pr}{\partial}
\newcommand{\hs}{\hspace{5mm}}

\newcommand{\dg}{\dagger}

\newcommand{\acc}{\\[3mm]}

\begin{document}
\begin{titlepage}
\strut\hfill
\vspace{0mm}
\begin{center}

{\large{ \bf  Lattice Three-Dimensional Skyrmions Revisited}}
\\[2mm]
{ E.G. Charalampidis${}^{1,2^{*}}$,  \ T.A. Ioannidou${}^{1^{\dg}}$ 
\ and \  P.G. Kevrekidis${}^{2^{\ddagger }}$}
\\[12mm]
$^1${\it School of Civil Engineering, Faculty of Engineering,
Aristotle University of Thessaloniki,\newline
 54124 Thessaloniki, Greece }\acc
$^2${\it  Department of Mathematics and Statistics, University of Massachusetts,  Amherst, \newline Massachusetts 01003-4515, USA}
\acc

\end{center}

\vspace{15mm}

\begin{abstract}
In the continuum a skyrmion is a topological nontrivial map between
Riemannian manifolds, and a stationary point of a particular energy
functional. This paper describes lattice analogues of the aforementioned
skyrmions, namely a natural way of using the  topological properties 
of the three-dimensional continuum Skyrme  model to achieve topological
stability on the lattice. In particular, using fixed point iterations,
numerically exact lattice skyrmions are  constructed;  and their stability
under small perturbations is explored by means of linear stability analysis. 
While stable branches of such solutions are identified, it is also
shown that they possess a particularly delicate bifurcation structure,
especially so in the vicinity of the continuum limit. The corresponding
bifurcation diagram is elucidated and a prescription for selecting
the branch asymptoting to the well-known continuum limit is given.
Finally, the robustness of the spectrally stable solutions is corroborated
by virtue of direct numerical simulations .
\end{abstract}

\vspace{15mm}

$^*${ Email: {\sf echarala@auth.gr}} \vspace{-2mm}

$^\dg${ Email: {\sf ti3@auth.gr}} \vspace{-2mm}

$^\ddagger${ Email: {\sf kevrekid@math.umass.edu}}

\end{titlepage}

\pagenumbering{gobble}
\clearpage
\pagenumbering{arabic}

\section{Introduction}
Lattice models involve discrete systems where the motion of a lattice 
particle depends on and also affects the motion of its neighbors. When
considering the long wavelength approximation of the discrete system, 
the latter is transformed into a continuum medium for which partial 
differential equations govern the local motion; see e.g. \cite{remoiss}-\cite{IPA}
for some case examples. Many attempts have been made in order to find 
methods that faithfully represent the continuum limits of discrete models
or vice versa that accurately capture the continuum phenomenology within 
the discrete case (the latter  is also relevant, in the context of numerical
computation). For a given continuum model, there are many different discrete
analogues which reduce to it in the continuum limit.

In \cite{SW} a scheme was proposed for generating one-dimensional topological 
{\it lattice} systems. It maintains an important feature of the continuum model,
namely the topological lower bound (so-called, Bogomolny bound) of the energy; 
while the lattice solitons saturate this lower bound. These are solutions of the
lattice Bogomolny equation. However, an open issue is whether  topology and 
topological stability of the solitons can be maintained on the lattice. In the
continuum, the stability of the topological solitons is often related to the
existence of such an energy bound but in the lattice,  it is not clear whether 
the corresponding  topological objects are also stable. The discretization scheme
of \cite{SW} has been applied in few continuum theories which have a Bogomolny 
bound \cite{L}-\cite{IK}; but also used in {\it non-topological} systems where 
a Bogomolny-type argument can be applied  \cite{IKV}. The aforementioned features
are not preserved in two-dimensional systems, since the energy of the lattice 
solitons is greater than the topological minimum \cite{Ward}.
 
In  this paper, following \cite{SW},  a lattice version of the  Skyrme model 
in $3+1$ dimensions is described based on a Bogomolny-type argument. To do so
we impose radial symmetry on the field, thus a one-dimensional system is  obtained,
and then its lattice version   which maintains a Bogomolny-type  bound is presented. 
A similar study, has been applied for studying the properties of  the $Q$-balls 
defined  at the continuum~\cite{IKV}. The corresponding  results were obtained 
analytically  and were close to the exact ones obtained by solving the full equations
numerically. 

The three-dimensional Skyrme model \cite{THR} serves as a popular model of the
dynamics of pions and nucleons, incorporating the former as its fundamental 
pseudo-Goldstone field and the latter in the form of topological solitons. Its
lattice  formulation is of some importance in its own right. The model is non-renormalizable
in perturbation theory and existing treatments of the model are semiclassical 
(quantizing only the  collective degrees of freedom of the soliton). A full 
quantization of the  theory requires a cutoff which can be attained by its 
lattice version.

\section{The Lattice Skyrme Model}
The Lagrangian density of the  Skyrme model is of the form
\be
12\pi^{2} {\cal L}=-\frac{1}{2}{\rm tr}\left(R_{\mu}R^{\mu}\right)
+\frac{1}{16}{\rm tr}\left(\left[R_{\mu},R_{\nu}\right]\left[R^{\mu},R^{\nu}\right]\right),
\label{lagrdens}
 \ee
where $R_\mu=\pr_\mu U U^{-1}$, the indices $\mu,\nu$ run from $0$ to $3$ 
and the metric is the Minkowski one, i.e.:  $g_{\mu\nu}=\mbox{diag}(1,-1,-1,-1)$
and $U$ is the $SU(N)$ Skyrme field. In order for finite energy field configurations
to exist,  the Skyrme field $U$  has to tend  to a constant matrix at spatial infinity. 
This compactifies the domain space of the Skyrme field $U$ into a large three-sphere 
and, thus, the static solutions are maps from $S^{3}\mapsto SU(N)$ and  as such can 
be classified by an integer-valued degree, (the so-called, \textit{topological charge} or 
\textit{baryon number}), denoted by $B$. Therefore a skyrmion is a finite energy field
configuration corresponding to topological solitons and carrying topological 
charge $B$.

In  \cite{HMS}, a separation of variables ansatz was introduced for the Skyrme field
by decomposing it into a radial and an azimuthal part. The former is captured by a
real profile function, namely $f(r)$ and the latter by a Hermitian projector $P$ that
depends only on the angular variables. 

Then, the kinetic and potential energy of the Skyrme model are equal to 
\bea
E_{\mbox{\tiny kin}}&=&\frac{1}{3\pi}\int\dot{f}^{2}\left(A_{N}r^{2}+2{\cal N} \sin^{2}{f}\right)dr, \nonumber \\
E_{\mbox{\tiny pot}}&=&\frac{1}{3\pi}\int\left(A_{N}r^{2}f_{r}^{2}+2{\cal N}\left(f_{r}^{2}+1\right)\sin^{2}{f}+{\cal I}\,\frac{\sin^{4}{f}}{r^{2}}\right)dr,\label{c}
\eea
respectively. 
Here, ${\cal N}$, ${\cal I}$ and $A_{N}$, are parameters independent of the
radial variable $r$, integrals of functions of  $P$ (for details, see \cite{IK} and
references therein); while $f(0)=\pi$ due to well posedness of (\ref{lagrdens}) and 
$f(\infty)=0$ due to  energy finiteness. The Skyrme model has no Bogomolny bound; 
however, a \textit{Bogomolny-type} argument leads to the bound
\bea
E_{\mbox{\tiny pot}}\!\!\!&\!\!\!=\!\!\!&\!\!\!\fr{1}{3\pi}\!\int \!\!\left\{\left( \sqrt{A_{N}}f_{r} r+\sqrt{\cal I}\fr{\sin^{2}{f}}{r}\right)^2\!\!+2{\cal N} \sin^{2}{f}(1+f_{r})^2   +2(2{\cal N}+\sqrt{A_N {\cal I}})\sin{f}\pr_r(\cos{f})\right\} dr\nonumber\\
&\geq& \fr{1}{3}(2{\cal N}+\sqrt{A_N {\cal I}}),
\label{B}
\eea
which is stronger than the usual Fadeev-Bogomolny one $E_{\mbox{\tiny pot}}\geq12\pi^2 {B}$.
Thus the discretization scheme of \cite{SW} can be applied in order to
examine whether the corresponding lattice skyrmions are stable. 

In \cite{IK}, based on such a Bogomolny-type argument, a discretization 
scheme was applied to derive spherically
symmetric three-dimensional skyrmions. However, an unfortunate sign error 
was present in the latter calculation that has been corrected herein.
In order to obtain an {\it appropriate} discrete version of
the Skyrme model, there are two critical points to be considered: 
a) the discretization formulae of the Skyrme fields and corresponding terms and
b) the choice of the discrete potential energy at the origin. In this note, 
both forms have been accounted for appropriately
in order to obtain a relevant result addressed below.

Hereafter, $r$ becomes a discrete variable with lattice spacing $h$ while the 
real-valued field $f(r,t)$ depends on the continuum variable $t$ and the 
variable $r=nh$ where  $n\in Z^+$. Then, $f_+=f\left((n+1)h,t\right)$  denotes 
a forward shift and thus, the forward difference is given by $\Delta f=(f_+-f)/h$. 
To obtain  a {\it lattice version of the Bogomolny-type bound},  we start with 
the last term of  (\ref{B}) and  discretize it as: $\sin f\pr_r(\cos f)=\sin f %
\Delta(\cos f)\equiv-\fr{2}{h}\sin f \sin \left(\fr{f_+-f}{h}\right) \sin \left(\fr{f_++f}{h}\right) $.
This leads to the following discretization choices:
\bea
f_r&\rightarrow&\fr{2}{h}\,\sin\left(\fr{f_+-f}{2}\right),\nonumber\\
\sin f&\rightarrow&\sin\left(\fr{f_++f}{2}\right).
\label{dfdis}
\eea
The same discretization scheme can be applied at the origin with the only 
assumption that at the origin the Bogomolny-type equation is satisfied, i.e. 
$\sqrt{A_{N}}f_{r} r+\sqrt{\cal I}\fr{\sin^{2}{f}}{r}\equiv0$.  Thus, the 
discrete kinetic and potential energy assume the form:
{\footnotesize \bea
E_{\mbox{\tiny kin}}\!\!\!\!&=&\!\!\!\!4\pi h \sum_{n=1}^{\infty}\left[A_N \,n^2h^2+2{\cal N}\sin^{2}\left(\fr{f_++f}{2}\right)\right] \dot{f}^2,\acc
E_{\mbox{\tiny pot}} \!\!\!\!&= &\!\!\!\!8\pi h\,\cos^2\left(\fr{f_1}{2}\right)\left[{\cal N}\left(\fr{4}{h^2}\,\cos^2\left(\fr{f_1}{2}\right)+1\right)+\fr{2}{h}\sqrt{A_N{\cal I}}\cos\left(\fr{f_1}{2}\right)\right]\nonumber\\
&+&\!\!\!\!4\pi h\sum_{n=1}^{\infty}\Bigg\lbrace 4A_{N}n^2\sin^{2}\left(\fr{f_+-f}{2}\right)+\fr{{\cal I}}{n^{2}h^{2}} \sin^4\left(\fr{f_++f}{2}\right)
 +2{\cal N}\left[\fr{4}{h^{2}}\sin^{2}\left(\fr{f_+-f}{2}\right)+1\right]\sin^2\left(\fr{f_++f}{2}\right)\Bigg\rbrace. \label{epot_dis}\nonumber\\
\eea}
Then, the Euler-Lagrange equations read
\begin{eqnarray}
&&\ddot{f}\left[A_{N}h^2+2{\cal N}\sin^{2}\left(\frac{f_{+}+f}{2}\right)\right]
+{\cal N}\dot{f}\left(\dot{f}_{+}+\dot{f}\right)\sin{\left(f_{+}+f\right)}-\frac{\cal N}{2}\dot{f}^{2}\sin\left(f_{+}+f\right)
\nonumber\\
&&=\frac{\sin{f_{1}}}{2 h^{2}}\left[3 h \sqrt{A_{N} {\cal I}} \cos{\left(\frac{f_{1}}{2}\right)}+{\cal N}\left(8\cos^{2}{\left(\frac{f_{1}}{2}\right)}+h^2\right)\right] \nonumber\\
&&+A_N \sin(f_+-f)-\frac{{\cal I}}{2h^{2}}\sin{\left(f_{+}+f\right)}\sin^{2}{\left(\frac{f_{+}+f}{2}\right)}   \nonumber \\
&&+\frac{{\cal N}}{2}\left[\frac{4}{h^{2}}\sin{\left(f_{+}-f\right)}\sin^{2}{\left(\frac{f_{+}+f}{2}\right)}-\sin{\left(f_{+}+f\right)}\left(\frac{4}{h^{2}}\sin^{2}{\left(\frac{f_{+}-f}{2}\right)}+1\right)\right], \hspace{5mm} n=1, \nonumber \\
&& \label{eor} \\
&&\ddot{f}\left[A_Nn^2h^2+2{\cal N}\sin^{2}\left(\frac{f_++f}{2}\right)\right]+{\cal N}\dot{f}\left(\dot{f}_{+}+\dot{f}\right)\sin{\left(f_{+}+f\right)}\nonumber\\
&&-\frac{{\cal N}}{2} \left[\dot{f}^{2}\sin{\left(f_{+}+f\right)}+\dot{f}_{-}^{2}\sin{\left(f+f_{-}\right)} \right]\nonumber\\
&&=A_Nn^2\sin(f_+-f)-A_N(n-1)^2\sin(f-f_-)\nonumber \\ 
&&+\frac{{\cal N}}{2}\left[\frac{4}{h^{2}}\sin{\left(f_{+}-f\right)}\sin^{2}\left(\frac{f_{+}+f}{2}\right)-\sin{\left(f_{+}+f\right)}\left(\frac{4}{h^{2}}\sin^{2}{\left(\frac{f_{+}-f}{2}\right)}+1\right)\right]\nonumber\\
&&-\frac{{\cal N}}{2}\left[\sin{\left(f+f_{-}\right)}\left(\frac{4}{h^{2}}\sin^{2}{\left(\frac{f-f_{-}}{2}\right)}+1\right)+\frac{4}{h^{2}}\sin{\left(f-f_{-}\right)}\sin^{2}\left(\frac{f+f_{-}}{2}\right)\right]\nonumber\\
&&-\frac{\cal I}{2h^2}\left[\frac{\sin{\left(f_{+}+f\right)}}{n^{2}}\sin^{2}{\left(\frac{f_{+}+f}{2}\right)}+\frac{\sin{\left(f+f_{-}\right)}}{\left(n-1\right)^{2}}\sin^{2}{\left(\frac{f+f_{-}}{2}\right)}\right], \hspace{5mm} n>1.\nonumber \\
\label{eom}
\end{eqnarray}
In what follows we consider the $SU(2)$ case (where the Skyrme field
$U$ is a mapping between three-spheres, since the $SU(2)$ group is isomorphic to
$S^{3}$) and focus on the $B=1$ topological sector. This implies (see, also, \cite{IK})
that the corresponding parameters take the values of $A_{N}=1$, ${\cal N}={\cal I}=1$,
respectively.

\section{Numerical Simulations}

In this section solutions of the Euler-Lagrange equations (\ref{eor})
and (\ref{eom}) are derived numerically by assuming a 
lattice consisting of $n_{\textrm{max}}=800$ nodes (unless explicitly
stated otherwise). In particular, a fixed point iteration is used to identify 
relevant steady states and linear stability analysis is applied  to study
their response to small perturbations. Although two 
principal (and disconnected between them) branches
of solutions are obtained and presented below, it turns out that only one 
tends smoothly towards the corresponding continuum limit of the system, 
as soon as $h\rightarrow 0$; see the relevant discussion below. 
Furthermore, the robustness of the solution segments
of both branches which are found to be linearly stable is corroborated by 
direct numerical simulations of the time evolution dynamics.

The investigation consists of three parts: {\it existence, stability} and
{\it dynamics}. Concerning the existence, the  numerical procedure used is
a Newton - Raphson algorithm  accompanied by  a suitable initial guess; in
this way a static solution is obtained (up to a prescribed tolerance)
after a few iterations. Note that at the origin (i.e., $n=0$), 
we impose explicitly the boundary condition $f(0)=\pi$ (i.e., we consider the
case of compactified $R^3$ and not that of the compactified $R^3$ minus a small
ball of radius $\delta$ around the origin, as the latter would not possess stable
topological solitons even in the continuum limit). At the right end of the 
computational domain (i.e., $n=n_{\textrm{max}}+1$), a homogeneous 
Dirichlet boundary condition is applied, enforcing the existence of a topological
lattice soliton.

\begin{figure}[!t]
\begin{center}
\vspace{-0.9cm}
\mbox{\hspace{0.2cm}
\subfigure[][]{\hspace{-1.0cm}
\includegraphics[height=.18\textheight, angle =0]{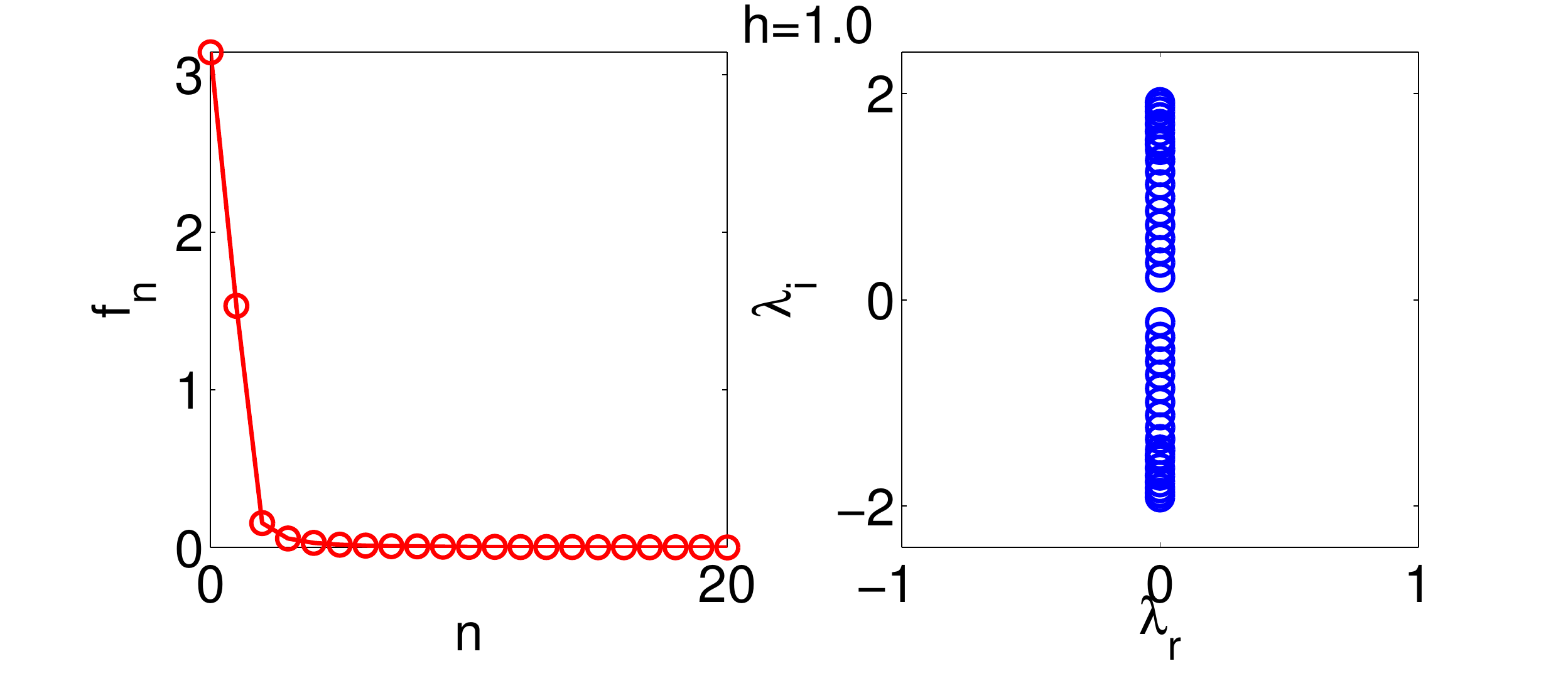}
\label{fig1a}
}
\subfigure[][]{\hspace{-1.0cm}
\includegraphics[height=.18\textheight, angle =0]{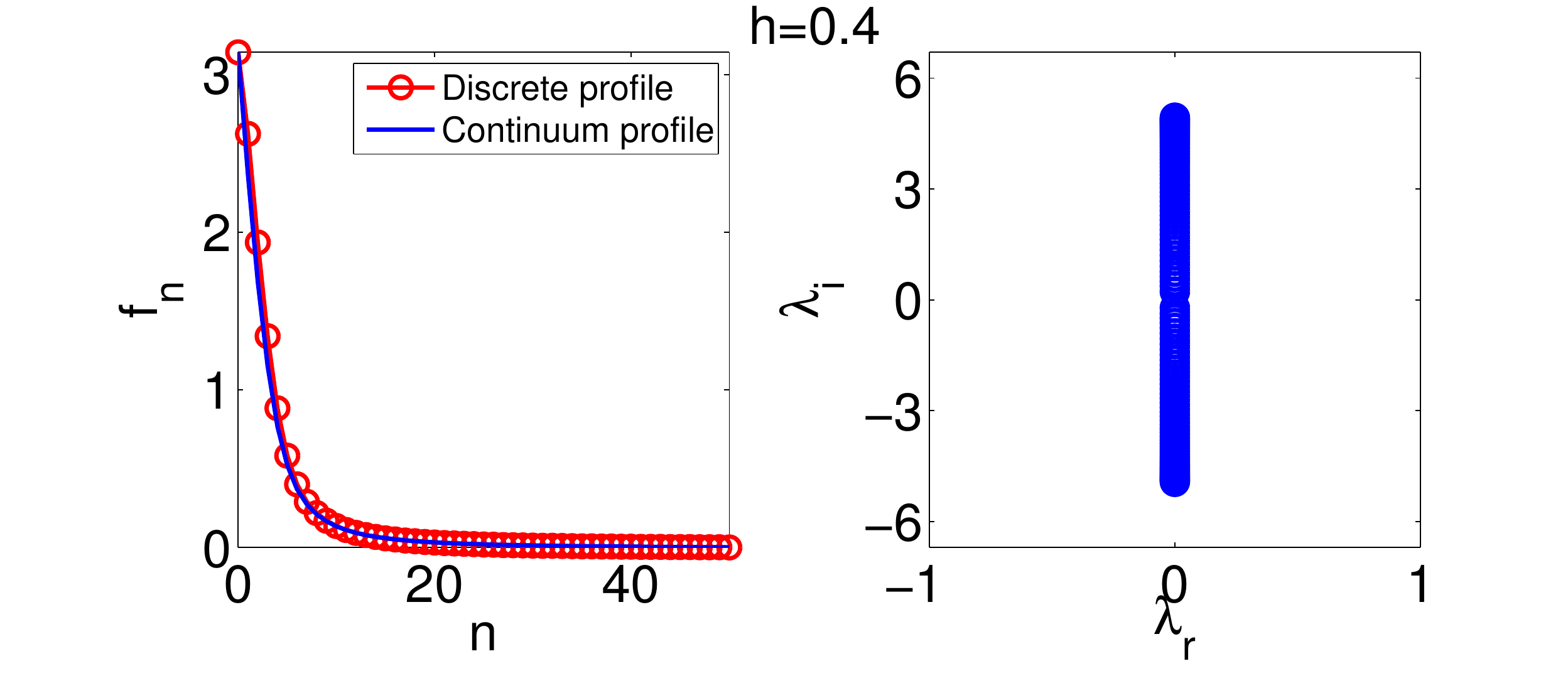}
\label{fig1b}
}
}
\mbox{\hspace{0.2cm}
\subfigure[][]{\hspace{-1.0cm}
\includegraphics[height=.18\textheight, angle =0]{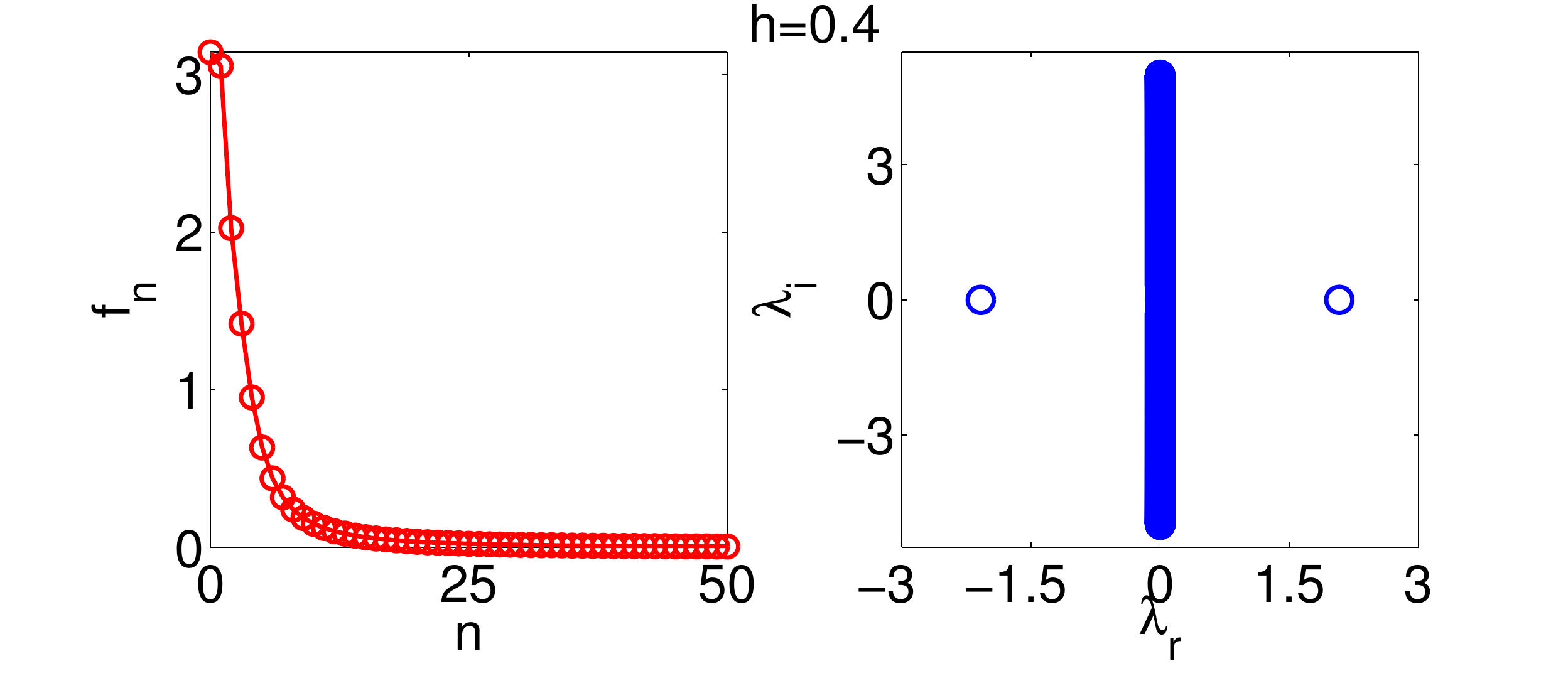}
\label{fig1c}
}
\subfigure[][]{\hspace{-1.0cm}
\includegraphics[height=.18\textheight, angle =0]{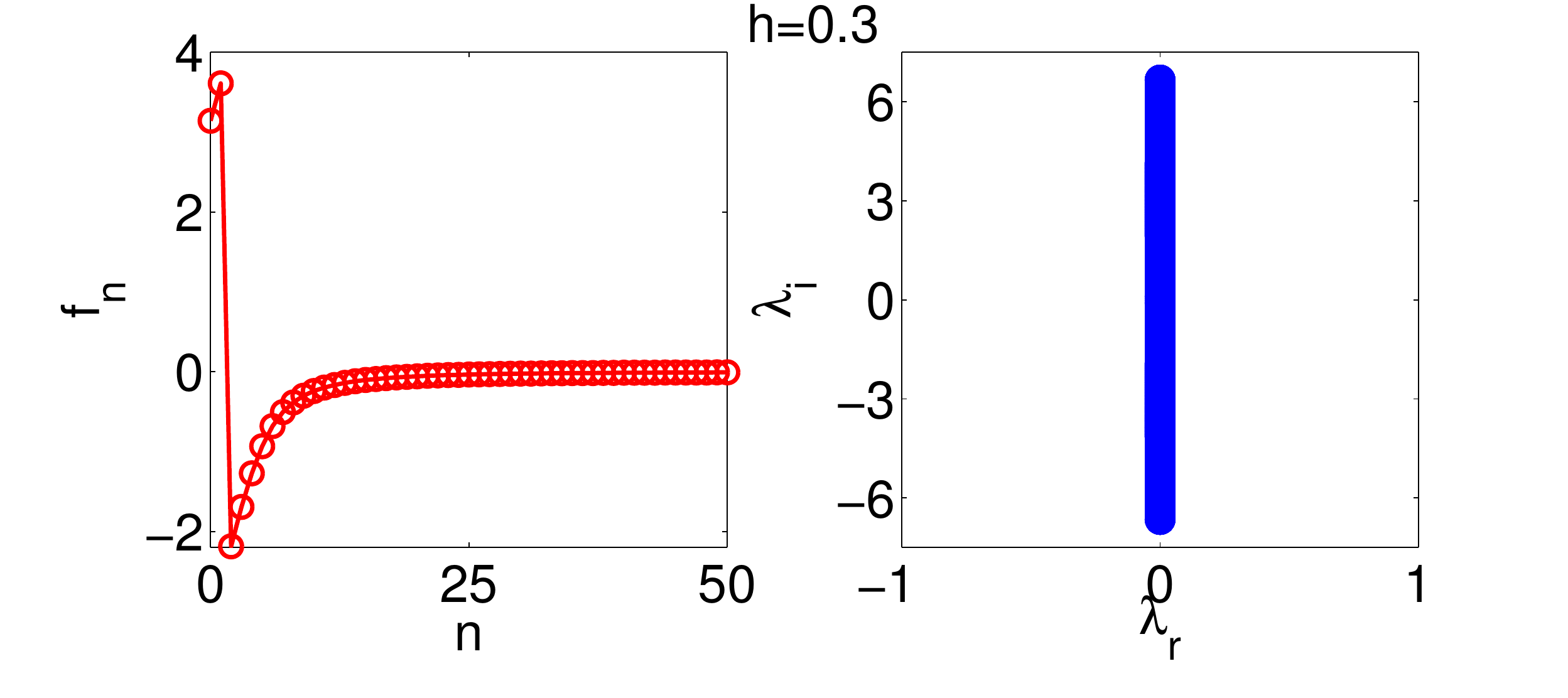}
\label{fig1d}
}
}
\mbox{\hspace{0.2cm}
\subfigure[][]{\hspace{-1.0cm}
\includegraphics[height=.18\textheight, angle =0]{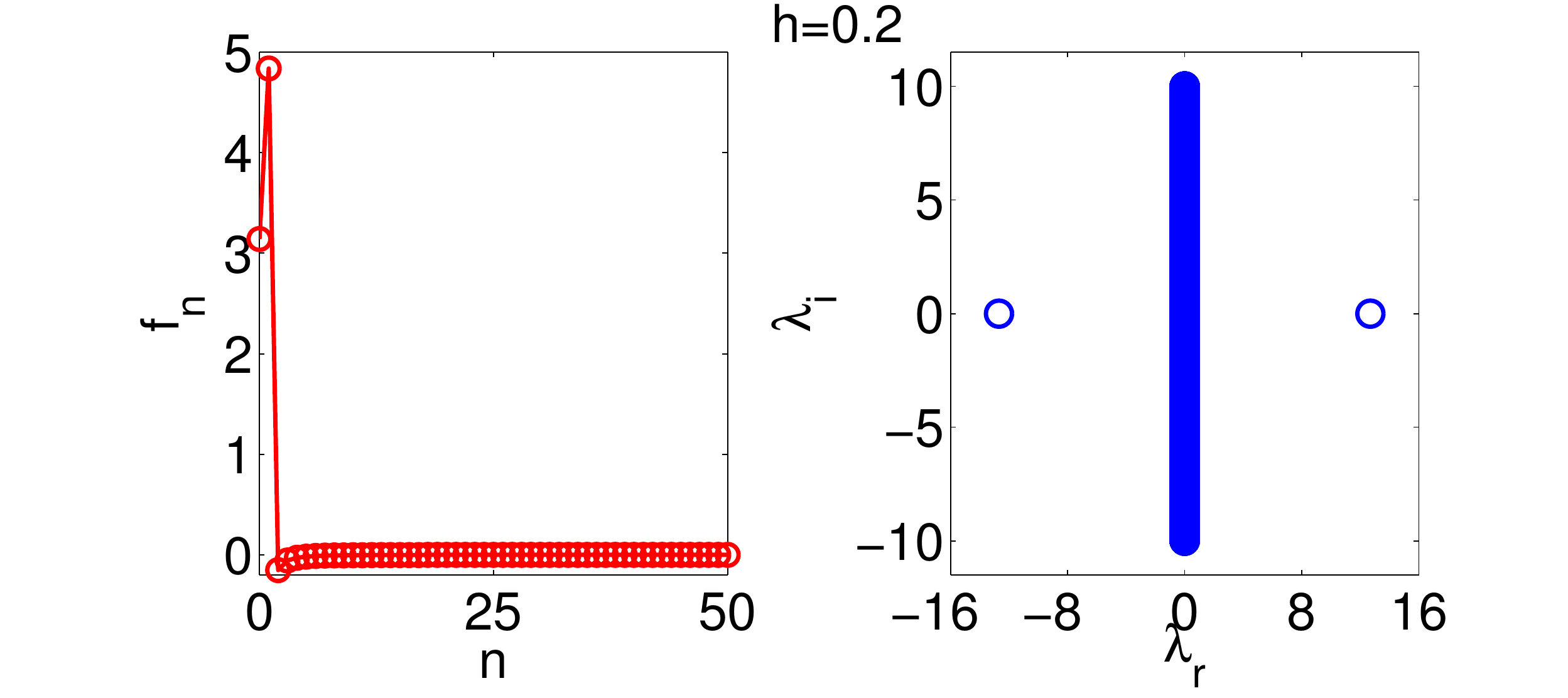}
\label{fig1e}
}
}
\end{center}
\vspace{-0.7cm}
\caption{(Color online) Static discrete profiles and corresponding 
eigenvalue spectra for the first branch of solutions are presented 
for a value of the lattice spacing of (a) $h=1$ (with $n_{\textrm{max}}=19$),
(b) $h=0.4$ (with $n_{\textrm{max}}=49$), (c) $h=0.4$, (d) $h=0.3$
and (e) $h=0.2$. These solutions correspond to the (a)-(e) labels 
of Fig.~\ref{fig3a}. In panel (b), the discrete (red circles)
against the continuum (solid blue line) profile is plotted for comparison.
}
\label{fig1}
\end{figure}

In Fig.~\ref{fig1}, the discrete profile function for various values of the
lattice spacing $h$ is shown. In particular, panel (a) corresponds to $h=1$ 
(an example of a rather {\it discrete} case) while panels (b)-(e) correspond
to $h=0.4, \ 0.4, \ 0.3$ and $0.2$ (examples gradually 
approaching the continuum case), 
respectively, and all labeled in connection with Fig.~\ref{fig3a}. As an initial
guess, a profile with  exponential decay of the form $f_{n}=\pi e^{-bnh}$ is
used, where $b$ is a width-controlling parameter. It is obvious from the top
panels of Fig.~\ref{fig1} that as $h$ decreases, the solution appears to
naturally and smoothly converge to its continuum counterpart;
nevertheless, as we will see below, this is {\it not} the 
branch that reaches the well-established continuum limit of the
model, due to the delicate structure of the bifurcation diagram
of this system. The latter limit is
illustrated in the top right panel of Fig.~\ref{fig1} with blue solid line as 
computed via a collocation method (see, for details, \cite{COLSYS}) applied to
the corresponding ordinary differential equation \cite{CIM}, thus yielding its
static radial topological soliton solution. Furthermore, 
although the profiles of panels (b)
and (c) correspond to the same value of $h$($=0.4$), they differ not only
structurally but the solution of panel (c) appears to be unstable according to
the
stability analysis that we will present next. Furthermore, we report the solutions
of panels (d) and (e) which appear structurally and physically not to be related with 
the skyrmion of the continuum limit. In particular, these solutions
are not only lacking the smoothness of the limit, but also the 
positive-definite structure of its profile. This is a 
feature that is not disallowed by our discretization, as $h \rightarrow 0$,
given the angular nature of our variables and the sinusoidal nature
of our associated discretization terms.

\begin{figure}[!ht]
\begin{center}
\vspace{-0.9cm}
\mbox{\hspace{-0.3cm}
\subfigure[][]{\hspace{-1.0cm}
\includegraphics[height=.18\textheight, angle =0]{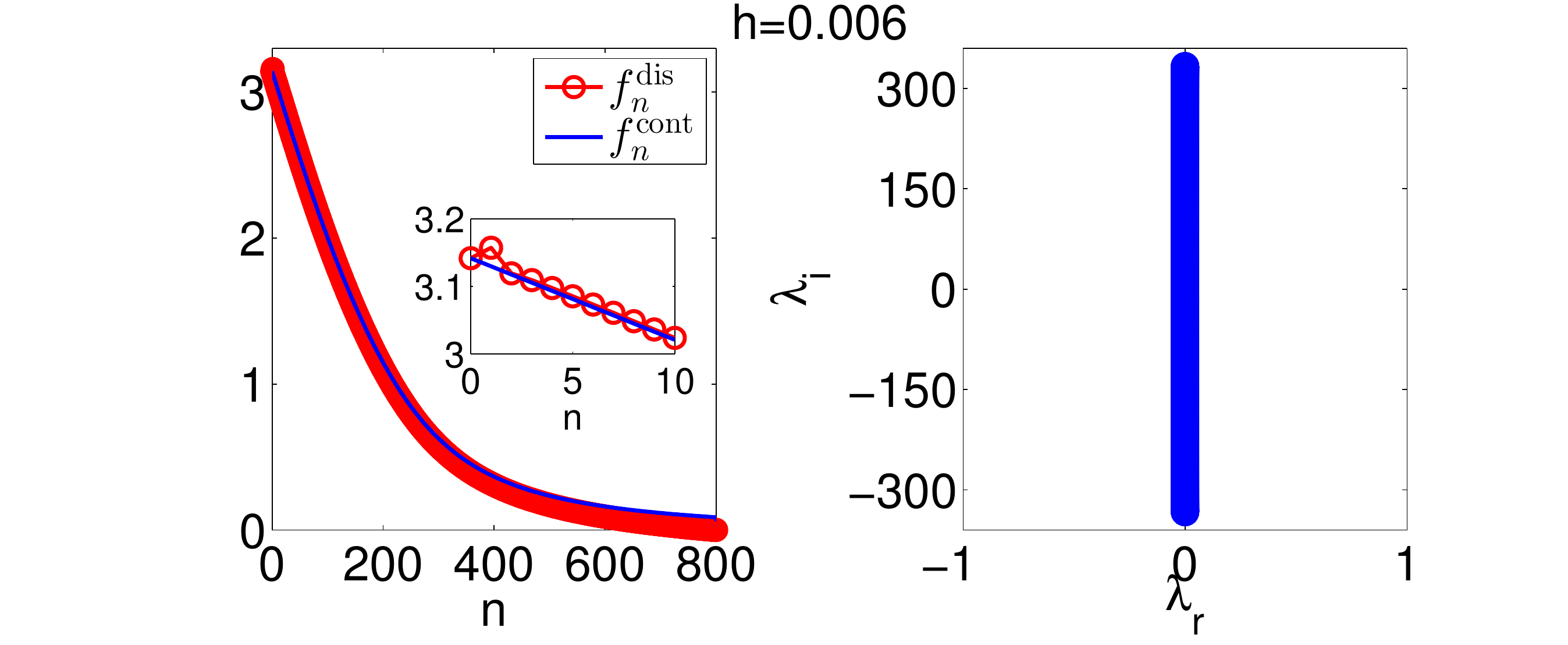}
\label{fig2a}
}
\subfigure[][]{\hspace{-1.3cm}
\includegraphics[height=.18\textheight, angle =0]{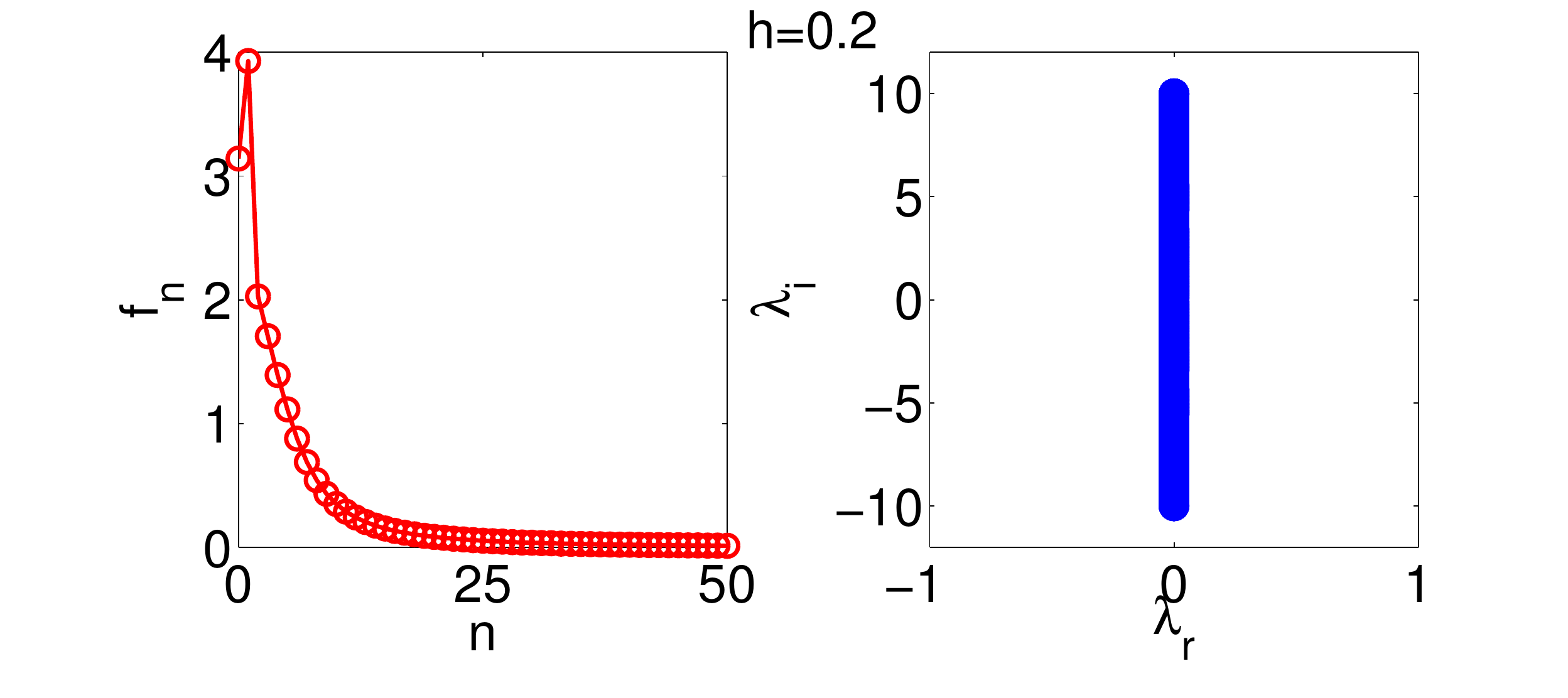}
\label{fig2b}
}
}
\mbox{\hspace{0.2cm}
\subfigure[][]{\hspace{-1.0cm}
\includegraphics[height=.18\textheight, angle =0]{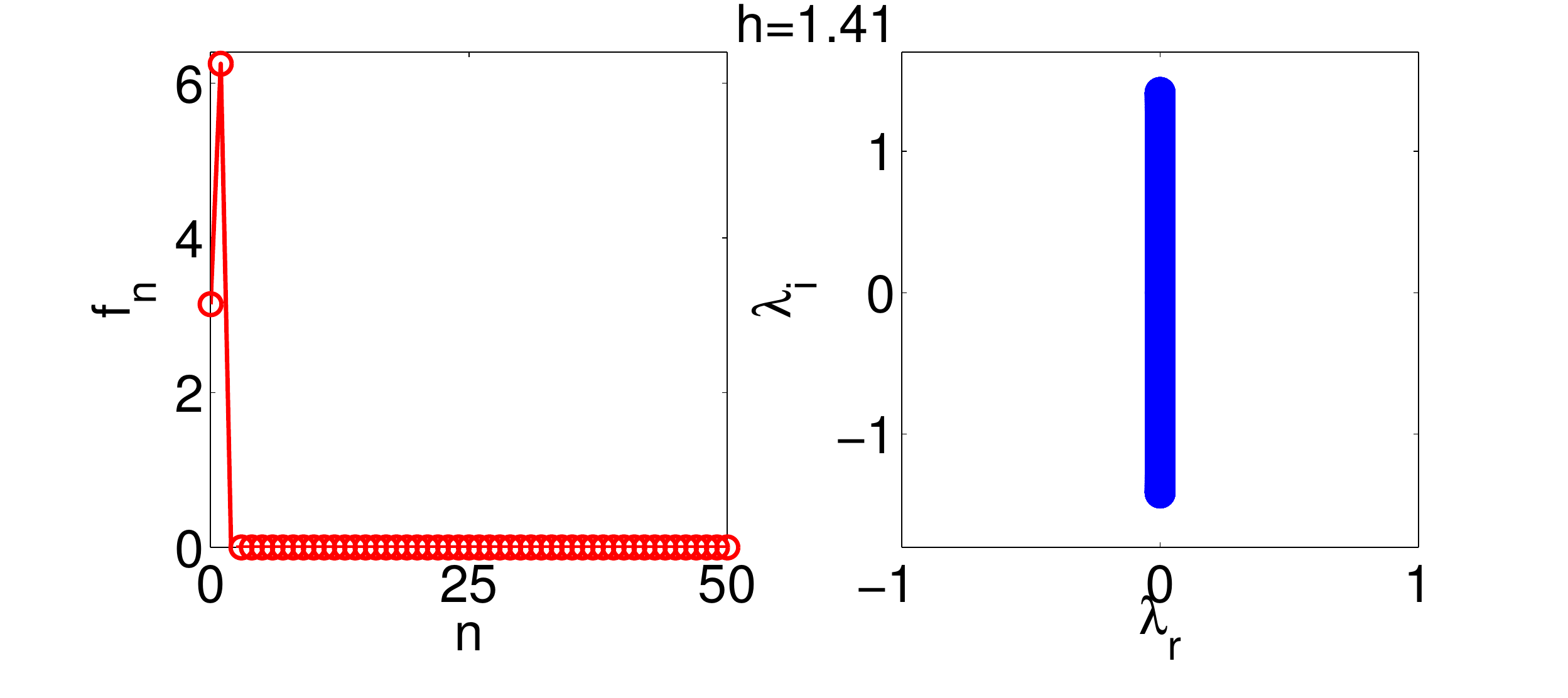}
\label{fig2c}
}
\subfigure[][]{\hspace{-1.0cm}
\includegraphics[height=.18\textheight, angle =0]{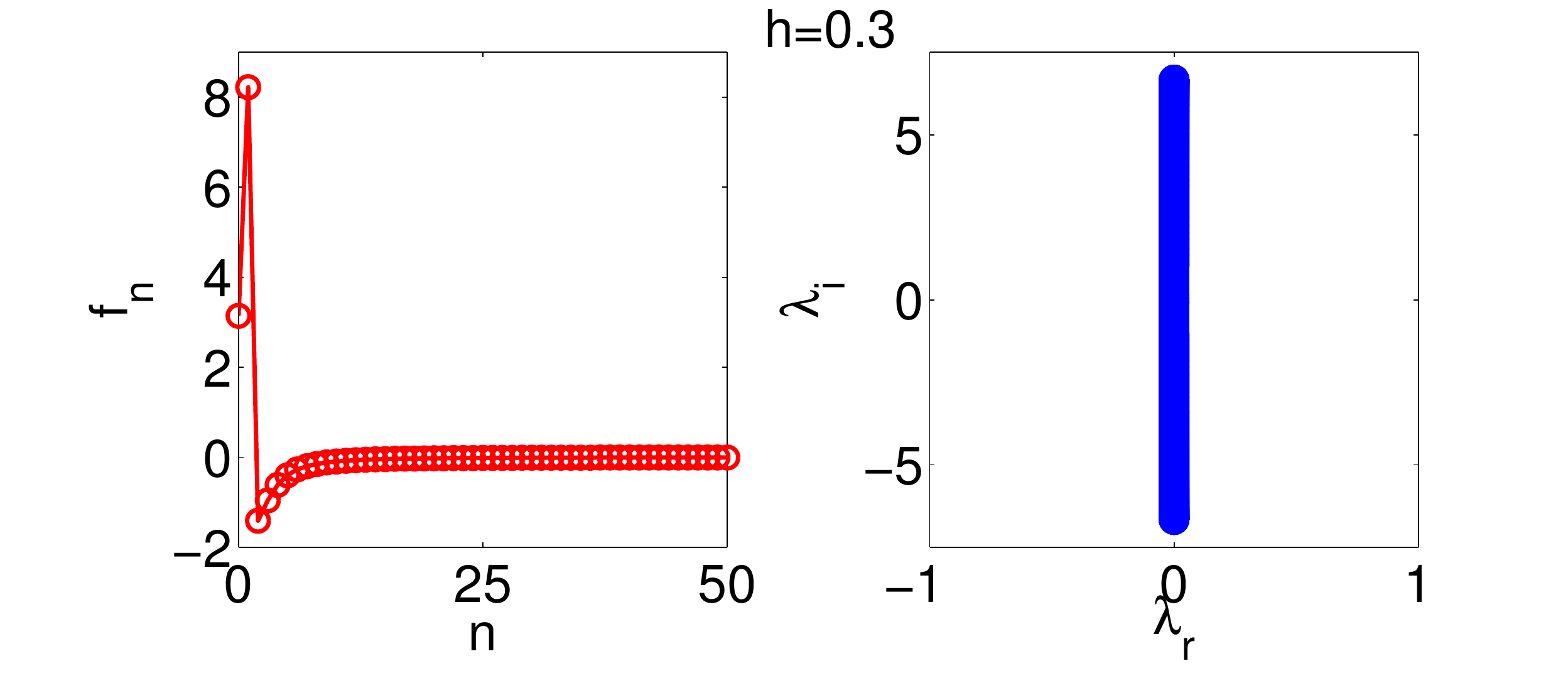}
\label{fig2d}
}
}
\end{center}
\vspace{-0.7cm}
\caption{(Color online) Same as Fig.~\ref{fig1} but for the second 
branch of solutions. Static discrete profiles and corresponding 
eigenvalue spectra for a value of the lattice spacing of (a) $h=6\times10^{-3}$, 
(b) $h=0.2$, (c) $h=1.41$ and (d) $h=0.3$. These solutions correspond
to the (a)-(d) labels of Fig.~\ref{fig4a}. In panel (a), the discrete (red circles)
against the continuum (solid blue line) profile is plotted for comparison.
}
\label{fig2}
\end{figure}

Concerning  the  stability of the obtained  lattice skyrmions the following
analysis is used. A linearization scheme around the stationary point $f^{0}$
is employed, in order to study the effect of  small perturbations. So, the 
profile function is chosen to be  of the  form:
\be
f_{n} = f_{n}^{0} + \epsilon \exp{\left(  \lambda t \right) }\,  w_{n},  \hs  (\epsilon \ll 1).
\ee

That way, at order $O(\epsilon)$, an eigenvalue problem is obtained with 
$\left(\lambda,w_{n}\right)$ representing  the  corresponding eigenvalue 
and eigenvector, respectively. Then, the eigenvalues $\lambda=\lambda_{r}+i \lambda_{i}$
without a  positive real part $\lambda_{r}$ correspond to oscillatory, 
marginally stable eigendirections (in our numerical computations presented
below, we consider a steady state solution having $\textrm{Max}(\lambda_{r})<5\times 10^{-4}$
to be stable). For instance, the eigenvalue spectra shown in panels (a),
(b) and (d) of Fig.~\ref{fig1} and panels (a)-(d) of Fig.~\ref{fig2} suggest 
that the corresponding solutions are dynamically stable since all the 
linearization eigenvalues are sitting on the imaginary axis. On the contrary,
if  the solution possesses either a real eigenvalue pair or a complex 
eigenvalue quartet (for the Hamiltonian problem at hand), this signals a
dynamical instability with a growth rate provided by the real part of the
corresponding eigenvalue. As it can be inferred from panels (c) and (e) of
Fig.~\ref{fig1}, the solutions turn out to be unstable characterized by a 
real eigenvalue pair.

Surprisingly, the solutions of Fig.~\ref{fig1} are {\it not} the only 
discrete solitons that exist. A second branch of discrete solitons exists
that was traced through the Newton-Raphson method. Four examples of them 
are shown in panels (a)-(d) of Fig.~\ref{fig2} for $h=0.006, \ 0.2,\  1.41$ and
$0.3$, respectively. This branch of solutions is found to be linearly 
stable and specifically, for very small values of $h$, matches very closely
the continuum limit of the topological soliton profile obtained through the
collocation method. 
Nevertheless, some comments are due here. 

\begin{figure}[!pt]
\begin{center}
\vspace{-0.9cm}
\mbox{\hspace{-0.5cm}
\subfigure[][]{\hspace{-1.0cm}
\includegraphics[height=.22\textheight, angle =0]{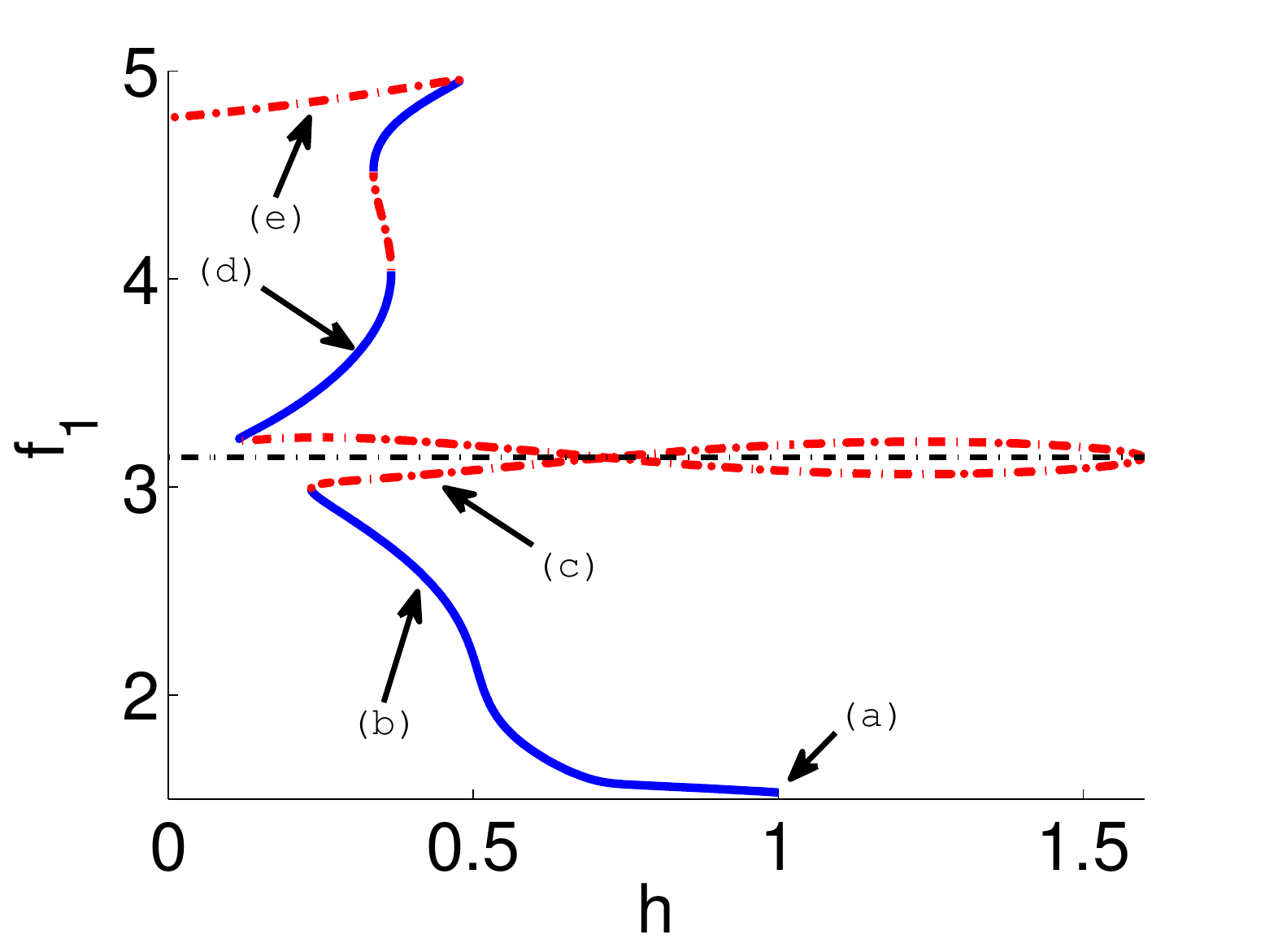}
\label{fig3a}
}
\subfigure[][]{\hspace{-0.5cm}
\includegraphics[height=.22\textheight, angle =0]{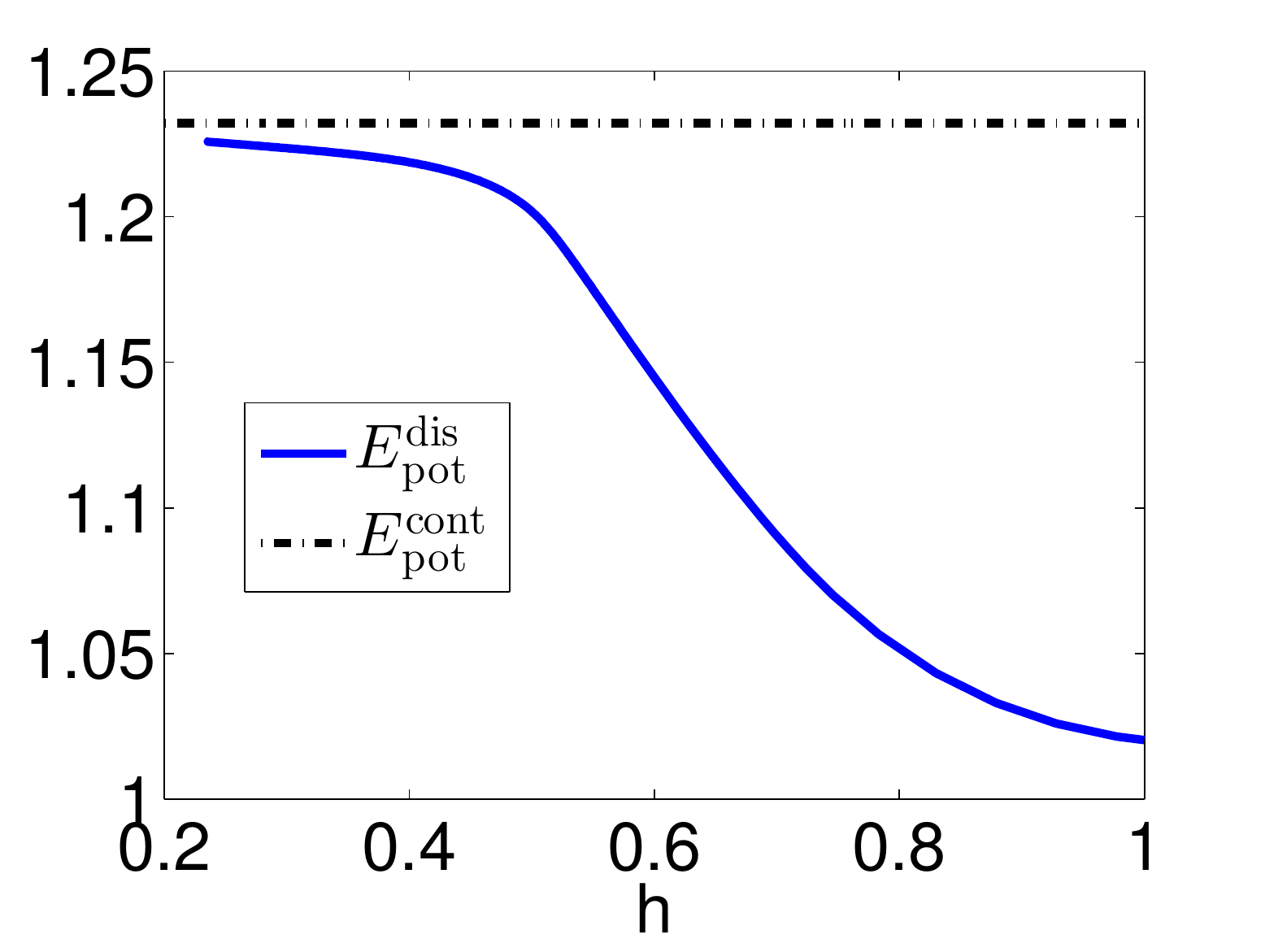}
\label{fig3b}
}
}
\end{center}
\vspace{-0.7cm}
\caption{(Color online) Results of continuation over the spacing parameter 
for the first branch of solutions: (a) Plot of the static discrete profile
at the first site of the domain as a function of the lattice spacing 
(i.e. $f_1(h)$). The solid (blue) segments correspond to stable regions while the 
dash-dotted (red) ones 
correspond to (real eigenvalue bearing) unstable parametric regions, respectively.
Note that the horizontal dash-dotted black line corresponds to the value of $\pi$.
(b) Plot of the normalized potential energy (i.e. $E_{\mbox{\tiny pot}}/12\pi^{2}$)
of our static solution (blue) as a function of $h$. The dash-dotted (black)
line corresponds to the actual value of the normalized continuum potential 
energy which is equal to 1.232.
}
\label{fig3}
\end{figure}
%

\begin{figure}[!t]
\begin{center}
\vspace{-0.9cm}
\mbox{\hspace{-0.5cm}
\subfigure[][]{\hspace{-0.3cm}
\includegraphics[height=.22\textheight, angle =0]{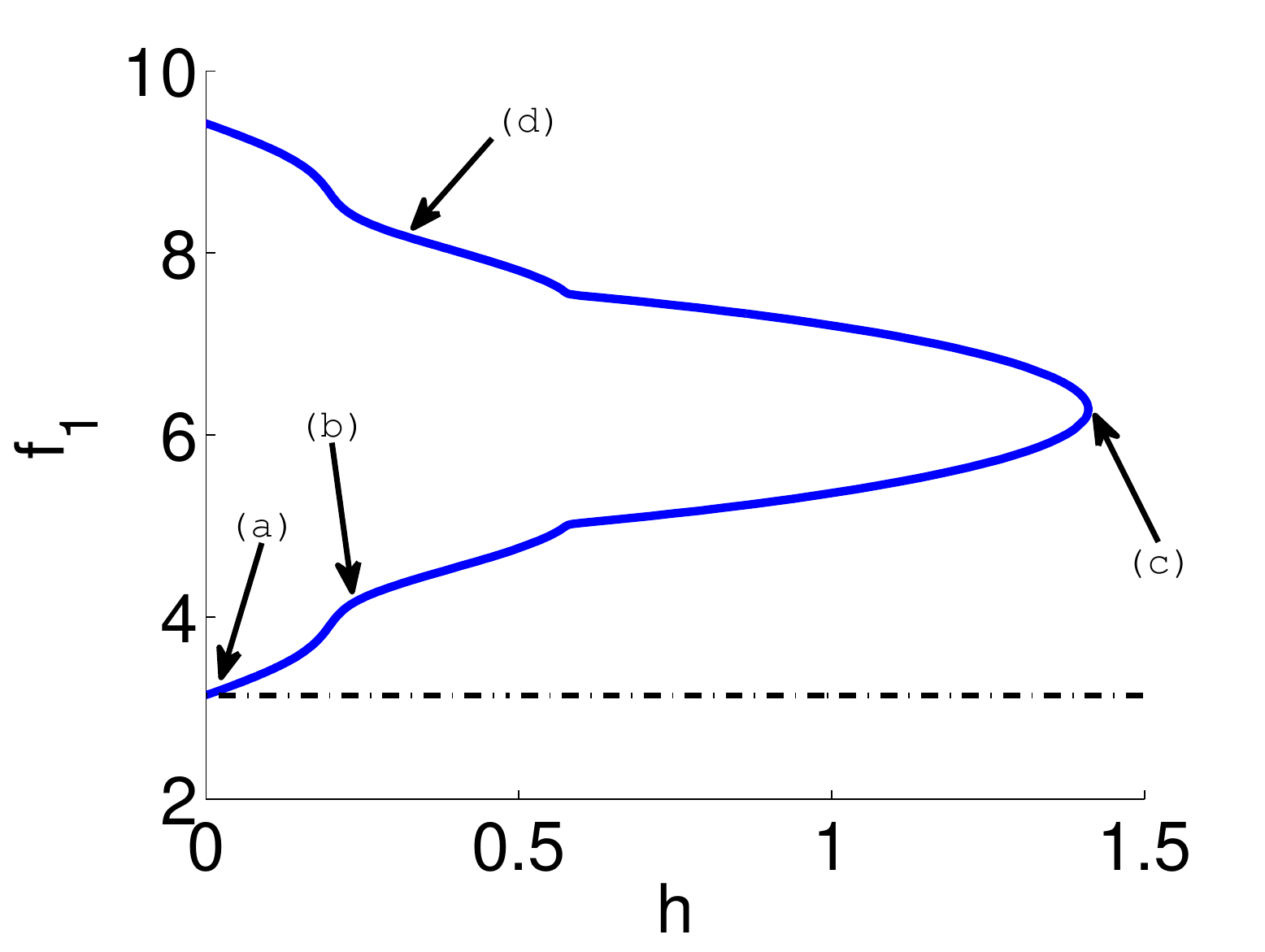}
\label{fig4a}
}
\subfigure[][]{\hspace{-0.3cm}
\includegraphics[height=.22\textheight, angle =0]{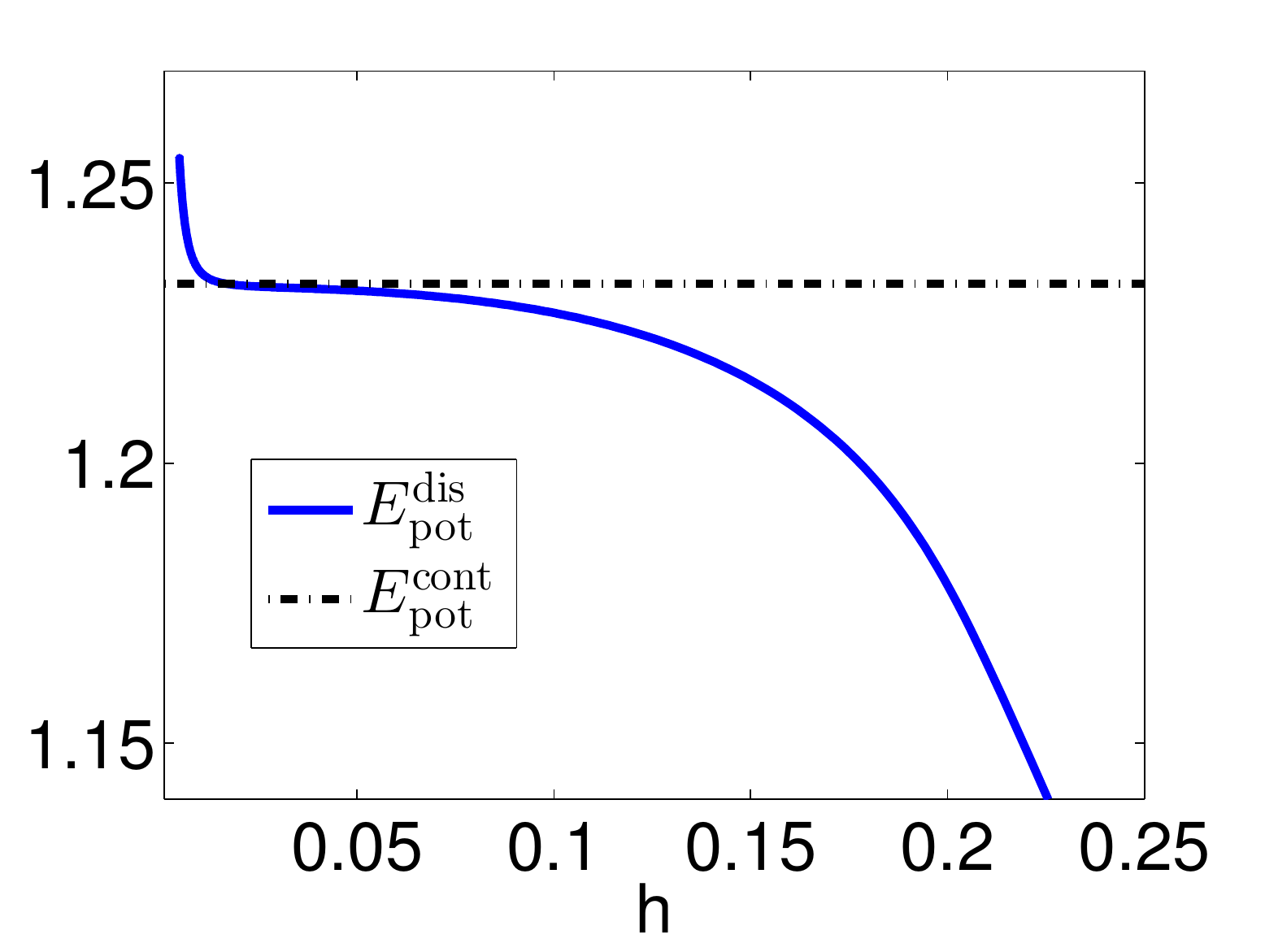}
\label{fig4b}
}
}
\mbox{\hspace{-0.5cm}
\subfigure[][]{\hspace{-0.3cm}
\includegraphics[height=.22\textheight, angle =0]{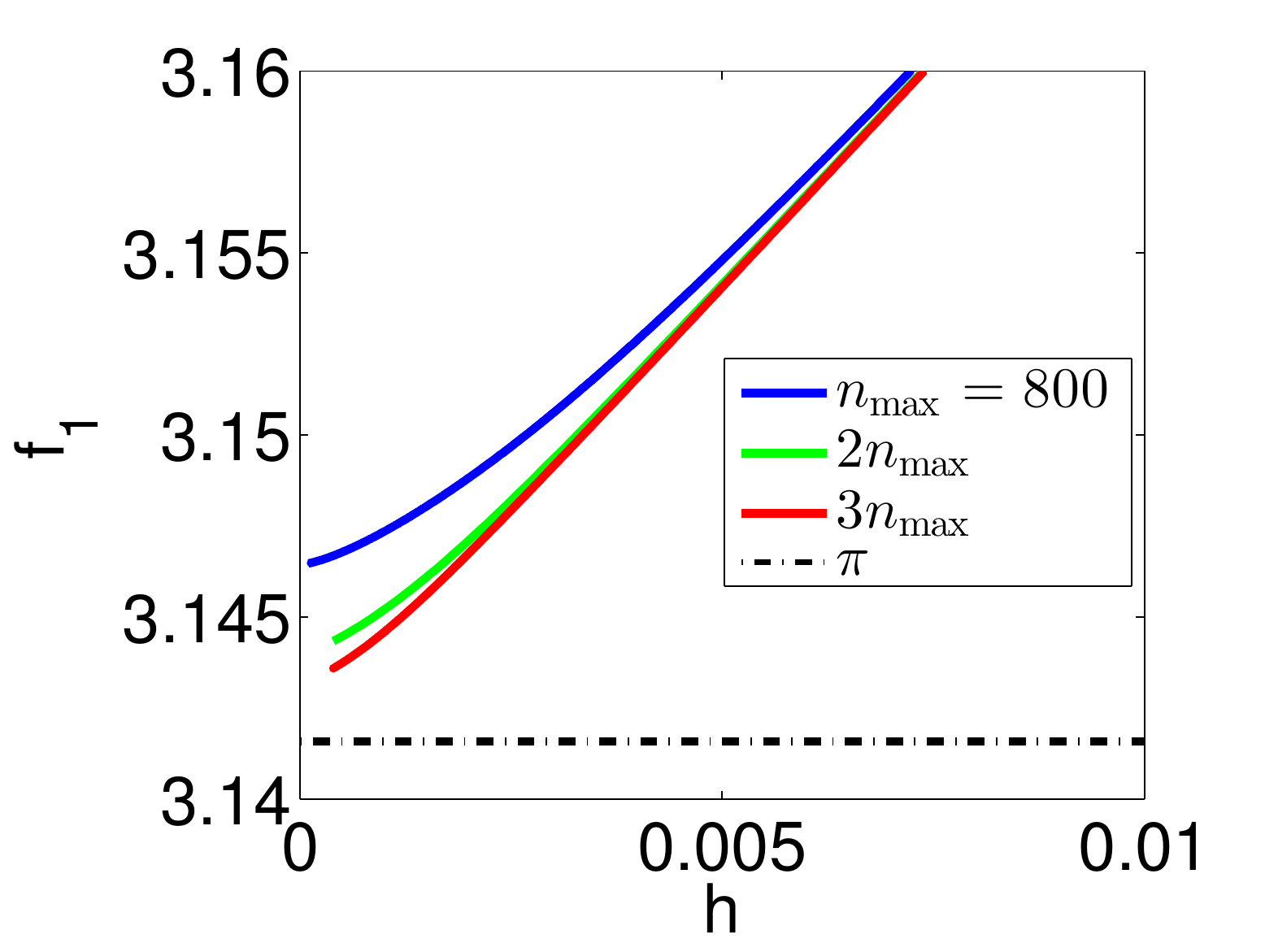}
\label{fig4c}
}
\subfigure[][]{\hspace{-0.3cm}
\includegraphics[height=.22\textheight, angle =0]{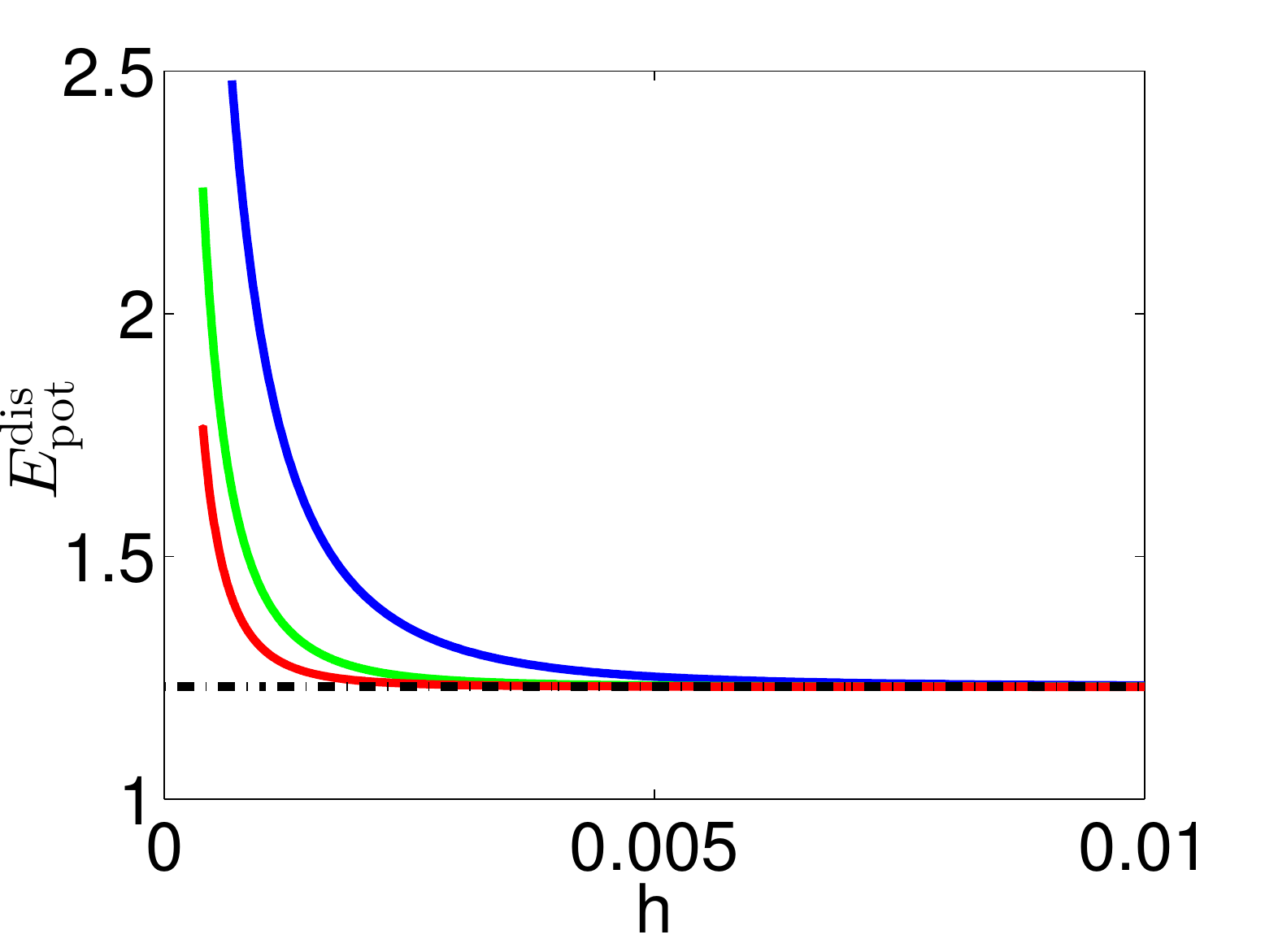}
\label{fig4d}
}
}
\end{center}
\vspace{-0.7cm}
\caption{(Color online) Same as Fig.~\ref{fig2} but for the second branch of 
solutions. Panels (c) and (d) are zoomed-in versions of (a) and (b) where
a direct comparison of the continuation results for values of $n_{\textrm{max}}=800$
(blue), $n_{\textrm{max}}=1600$ (green) and $n_{\textrm{max}}=2400$ (red), 
respectively, is shown.
}
\label{fig4}
\end{figure}

Although this second branch of solutions appears to very accurately capture
the continuum limit, the inset of Fig.~\ref{fig2a} suggests a very slight
mismatch near the origin (more precisely, at $n=1$). Admittedly, this disparity
must be introduced by our special way of handling the $n=1$ site to avoid
singularities in the discrete setting (cf. equation (\ref{eom})). 
In order to
understand the differences between these two branches, the characteristics of
continuation of the obtained solutions over the lattice spacing $h$ (using 
the computer software AUTO \cite{doedel_I,AUTO}) are displayed in Fig.~\ref{fig3}
and the top row of Fig.~\ref{fig4}, in terms of the value of the profile $f_{n}$
at $n=1$ (see, panels (a)) together with the corresponding total potential energy
$E_{\textrm{pot}}$ (see, panels (b)) given by Eq.~(\ref{epot_dis}). These parametric
continuations start from a large value of $h$ and, through progressive 
small decrements, tending towards the continuum limit of $h\rightarrow 0$, 
illustrate the distinction between the branches. However, when they
encounter turning points, the pseudo-arclength nature of the continuation
enables the code to bypass the relevant folds in the bifurcation diagram,
allowing the visualization of the complex bifurcation pattern.


The first branch of Fig.~\ref{fig3} is initiated at the value of the lattice 
spacing of $h=1$ (see, Fig.~\ref{fig1a} and label (a) in Fig.~\ref{fig3a}) 
and smoothly
approaches the continuum limit as $h$ is decreased (see, Fig.~\ref{fig1b} and 
label (b) in Fig.~\ref{fig3a}), as shown for the ordinate of the first site 
and also,
for the potential energy of the solution depicted in Fig.~\ref{fig3b}. 
However, this
first branch presents an intriguing feature: 
although it appears to asymptote smoothly towards the proper continuum counterpart, 
a saddle-node bifurcation at $h\approx0.234$ appears, followed by a turning point 
(at $h\approx1.597$) and an eventual cascade of saddle-node bifurcations (starting
at $h\approx0.115$), changing structurally the discrete profiles and the stability thereof. 
For the latter, it should be pointed out that stable
and unstable regions are presented in Fig.~\ref{fig3a} with solid blue and 
dash-dotted red lines, respectively. 
Furthermore, the solution branch labeled with (e) in Fig.~\ref{fig3a} (see also panel 
(e) in Fig.~\ref{fig1}) can be continued down for very small values of $h$, 
although the obtained solutions are unstable while the ordinate of the first
site does not asymptote to the correct continuum limit. The lack
of smoothness (a priori not disallowed by our discretization, as
per the discussion above) and the lack of stability suggest that this
continuation does not asymptote to the proper continuum limit solution.
Nevertheless, this is remedied by the second branch of solutions that
we now consider.
%

%
%

Indeed, it was this inability to identify the correct continuum limit as  
$h\rightarrow 0$ that led us to the second branch of solutions presented in 
Figs.~\ref{fig2} and \ref{fig4}. This branch is {\it not} of the 
monotonic (profile) type as was the case in some parametric regions of the first
one, although it is stable throughout the range of the lattice spacing $h$
considered. As 
shown in the top left panel of Fig.~\ref{fig4}, the profile function of the first
site is always $f_1 > \pi$ (see also, panels (b)-(d) in Fig.~\ref{fig2}) while
the bottom segment labeled by (a) only approaches $\pi$ as $h\rightarrow 0$. This feature
appears to be less physical and a particular 
attribute of the discretization imposed
here for $n=1$.  However, it can be smoothly continued to extremely small values
of $h$ thus, approaching the corresponding continuum limit (as shown with the solid
blue line in Fig.~\ref{fig4c}). In particular, at such small values of $h$ the precise (distinct
from $n \neq 1$) form of the discrete equation for $n=1$ appears to be 
responsible for
the induced {\it jump} (see the inset of Fig.~\ref{fig2a}) and also for the overshooting
evident in the top right panel of Fig.~\ref{fig4}. To investigate
this issue further and confirm the approach of the proper
continuum limit by the lower portion of this branch, we increased 
the total number of the nodes $n_{\textrm{max}}$ in the 
one-dimensional lattice. The bottom row of Fig.~\ref{fig4} summarizes our findings
using $n_{\textrm{max}}=800, 1600$ and $2400$ lattice nodes where panels (c) and (d)
correspond to the zoom-ins of panels (a) and (b), respectively. It can be discerned
from Fig.~\ref{fig4c} that the trend of $f_{1}(h)$ gradually asymptotes to the correct
continuum limit as the total number of the nodes is increased. The latter suggests
that for a sufficiently ``large'' lattice we are able to recover the correct continuum
radial profile, thus establishing the consistency of the discretization scheme (\ref{dfdis})
employed. Finally, the overshooting that appears 
in the discrete potential energy in 
Fig.~\ref{fig4b} gradually disappears as shown in Fig.~\ref{fig4d} for larger values
of $n_{\textrm{max}}$, therefore enforcing the validity of the scheme proposed.
Hence, clearly the lower portion of the second branch not only is smooth
but also approaches a smooth, positive monotonic profile in the continuum
limit, thereby properly asymptoting to the continuum soliton. However,
it should be noted that this branch of solutions too, as $h$ is increased
goes around a turning point; this implies that this family
of solutions cannot be continued indefinitely within the highly discrete regime.
Past this fold, the solution again acquires a non-smooth, non-sign-definite
form, hence not being a suitable candidate for continuation towards
the continuum limit. It is thus through the smoothness (and, in part,
the positive-definiteness and the stability properties) of the solution
that we select the suitable candidate (among the different available
solutions for small $h$) for reaching the continuum limit. 

\begin{figure}[!t]
\begin{center}
\vspace{-2.0cm}
\mbox{\hspace{-0.5cm}
\subfigure[][]{\hspace{-0.2cm}
\includegraphics[height=.24\textheight, angle =0]{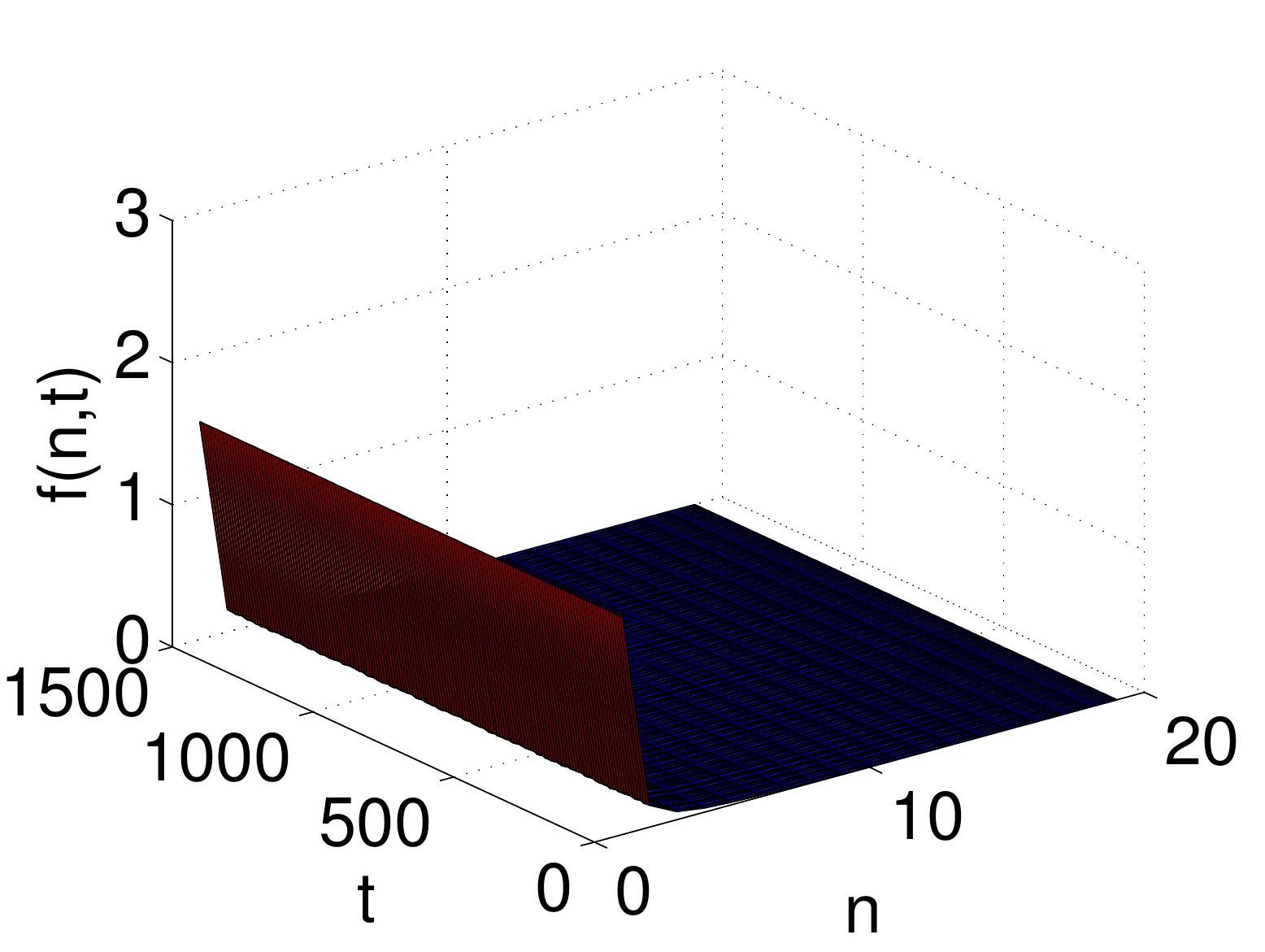}
\label{fig5a}
}
\subfigure[][]{\hspace{-0.2cm}
\includegraphics[height=.24\textheight, angle =0]{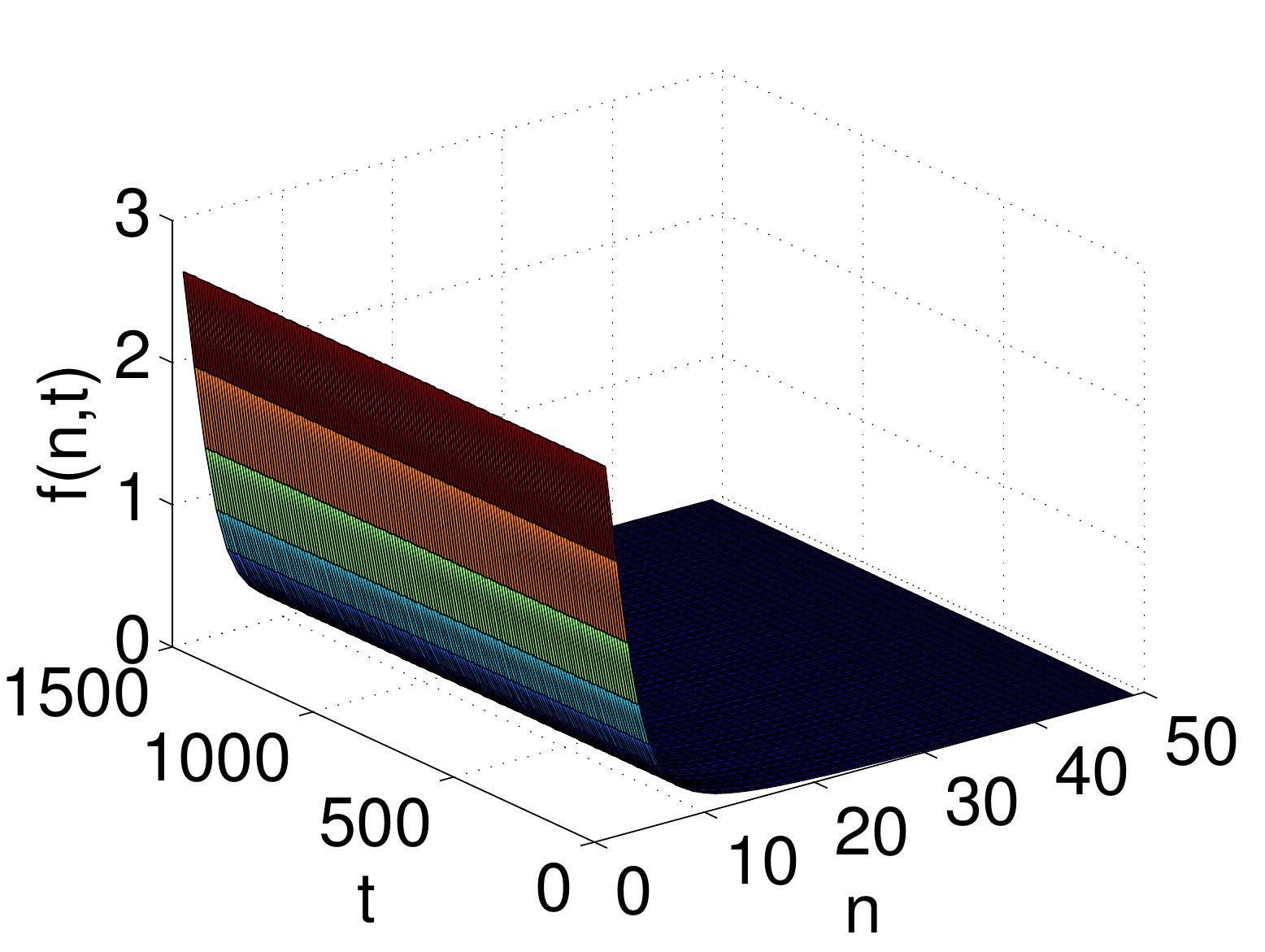}
\label{fig5b}
}
}
\mbox{\hspace{-0.5cm}
\subfigure[][]{\hspace{-0.2cm}
\includegraphics[height=.24\textheight, angle =0]{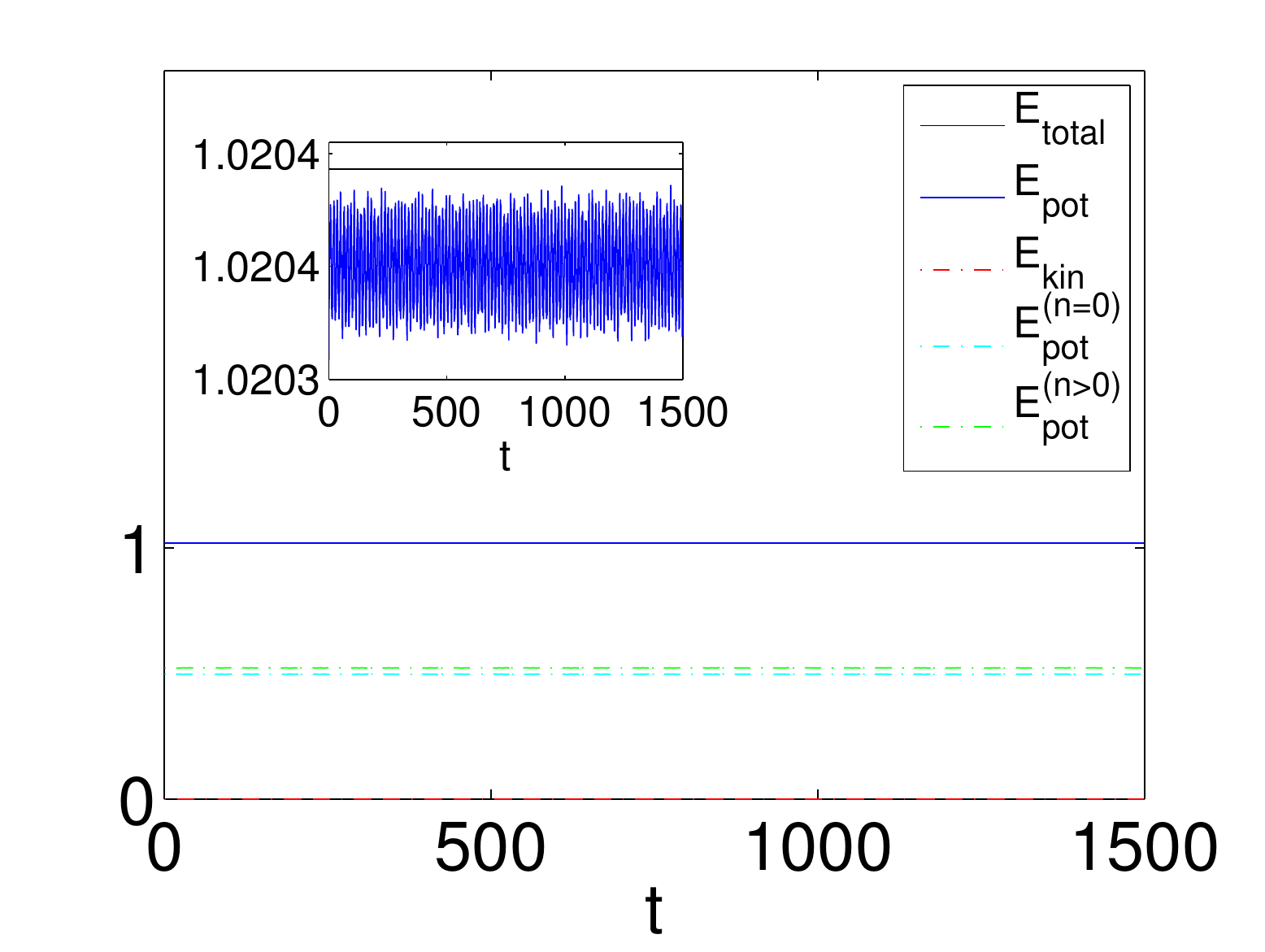}
\label{fig5c}
}
\subfigure[][]{\hspace{-0.2cm}
\includegraphics[height=.24\textheight, angle =0]{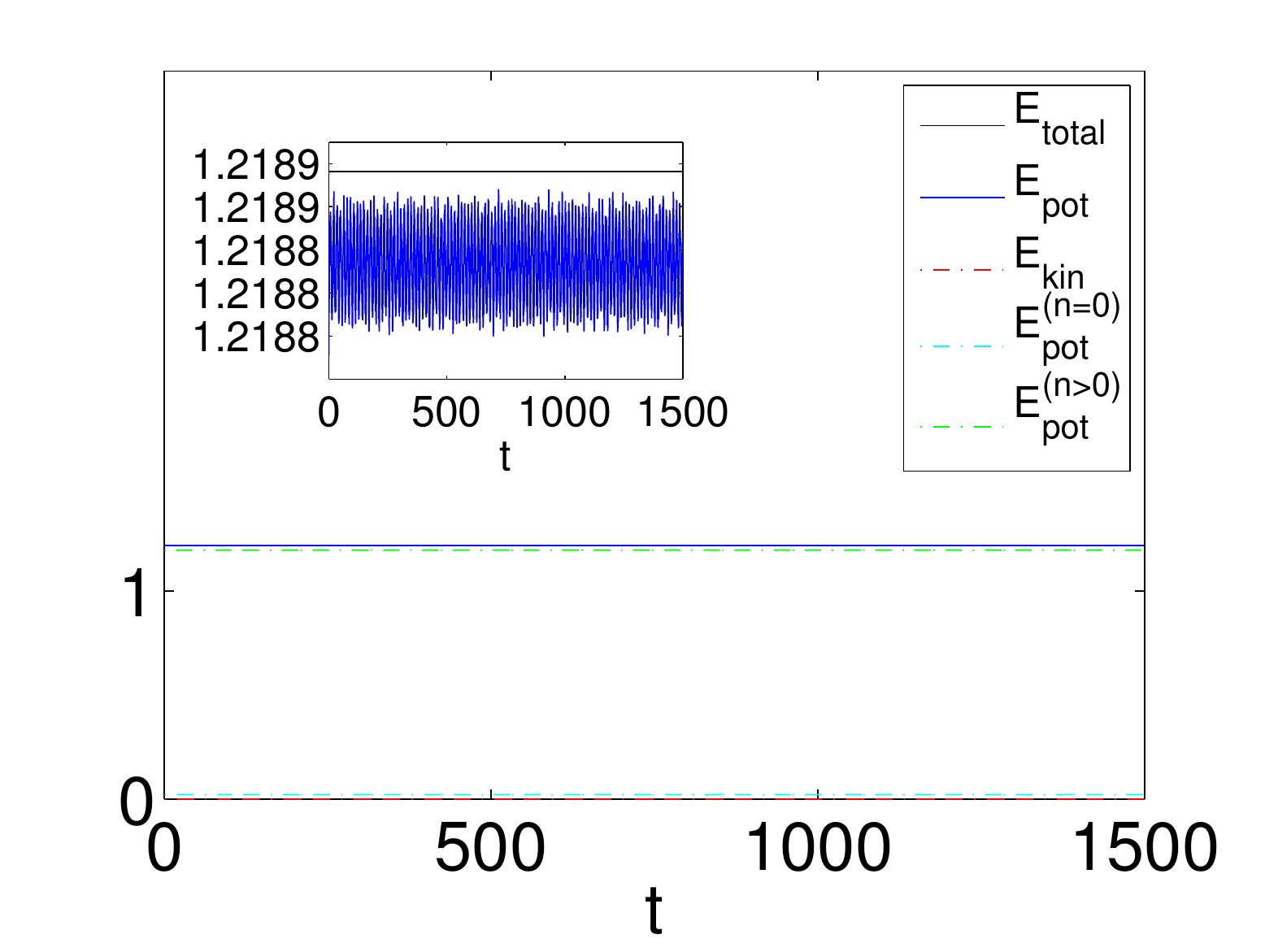}
\label{fig5d}
}
}
\mbox{\hspace{-0.5cm}
\subfigure[][]{\hspace{-0.2cm}
\includegraphics[height=.245\textheight, angle =0]{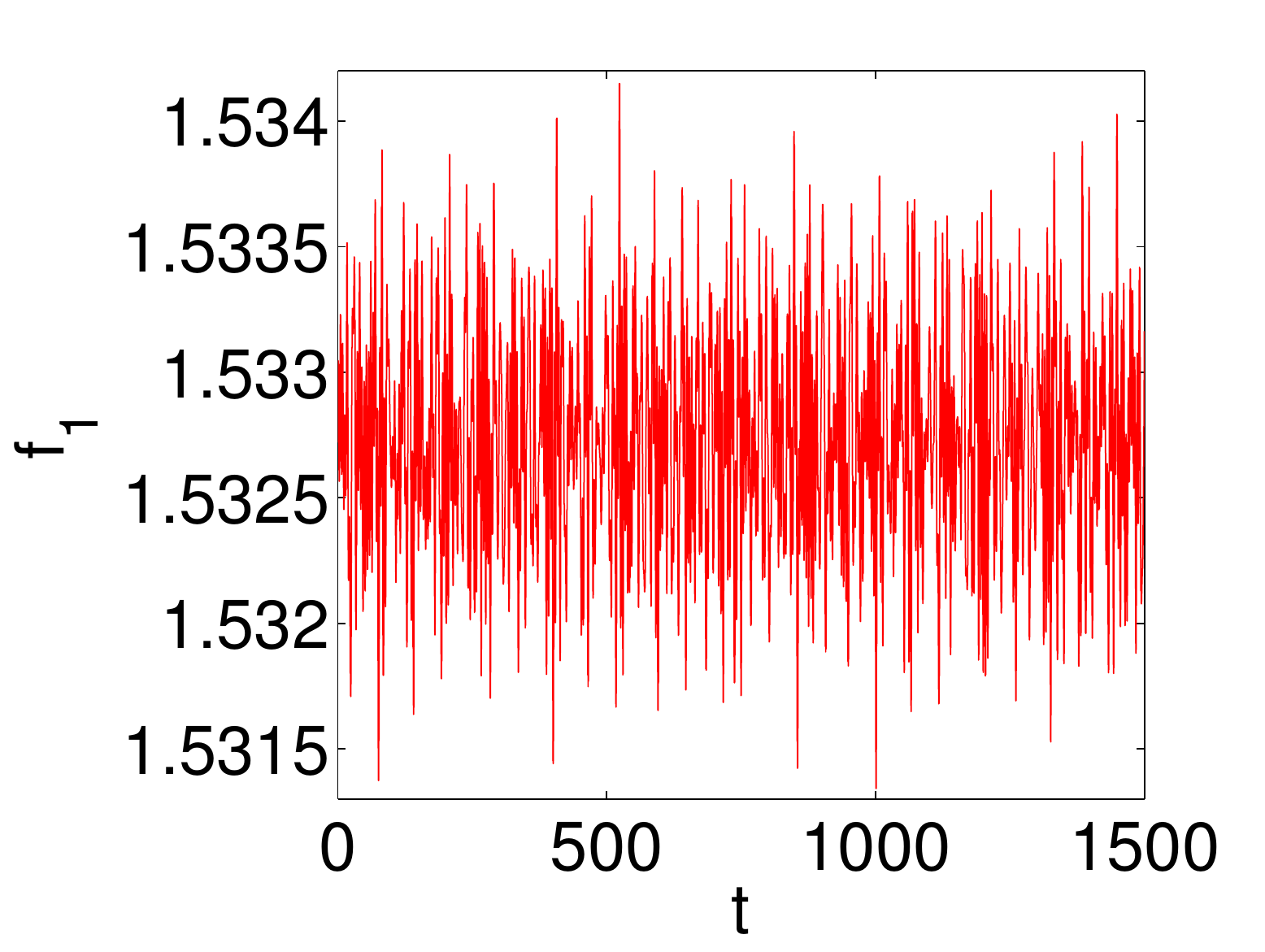}
\label{fig5e}
}
\subfigure[][]{\hspace{-0.2cm}
\includegraphics[height=.245\textheight, angle =0]{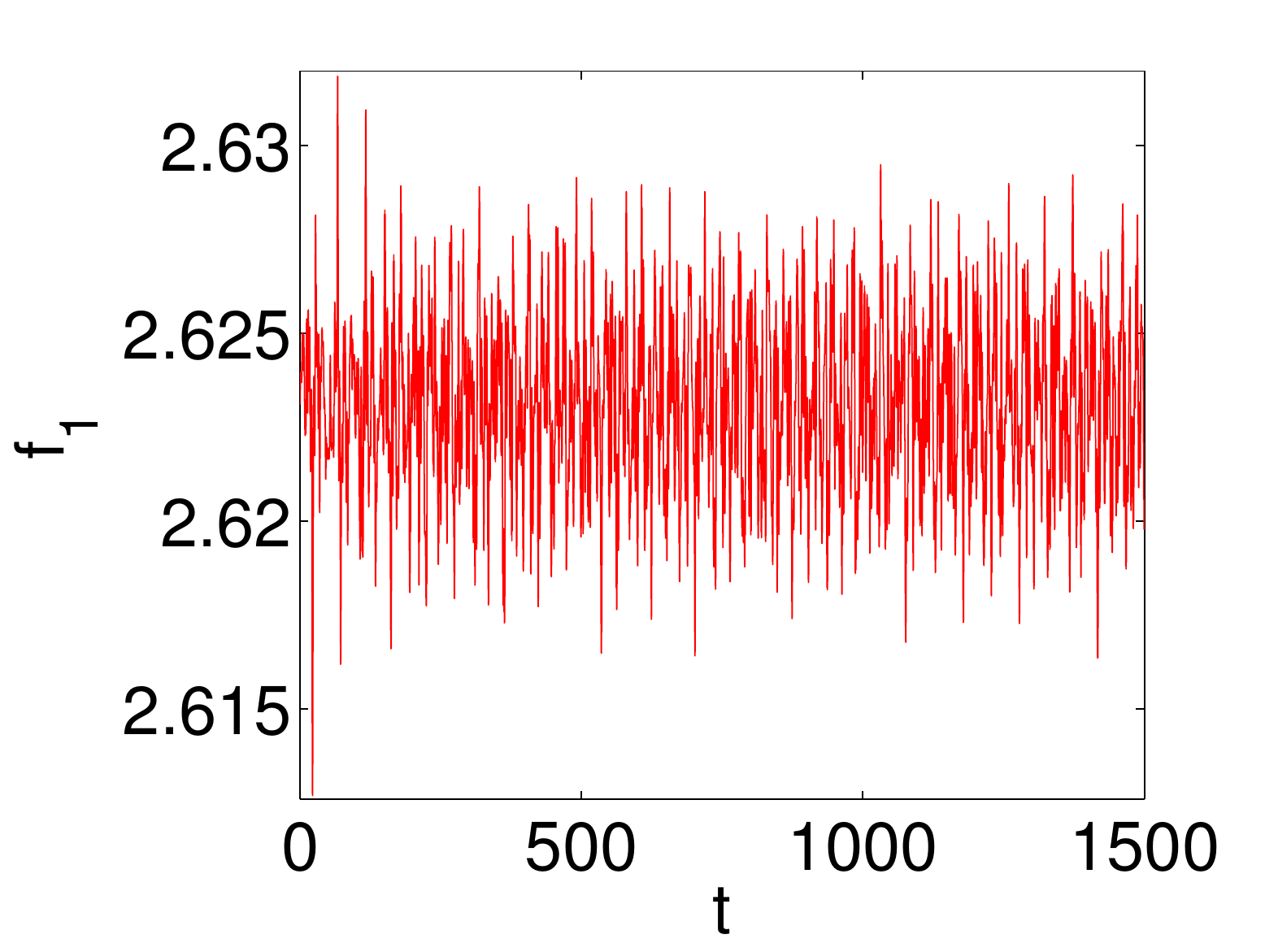}
\label{fig5f}
}
}
\end{center}
\caption{(Color online) Time evolution of the perturbed solutions of 
Fig.~\ref{fig1}. Left and right panels correspond to lattice spacing
$h=1$ and $h=0.4$, respectively. The top panels (a-b) show the space-time
evolution of the profile function $f(n,t)$. The middle panels (c-d) illustrate
the evolution of the kinetic and potential energies, and the conservation of
the total energy. The bottom panels (e-f) show the dependence of the profile 
function $f_1$ versus time, upon the imposition of the small random perturbation.
}
\label{fig5}
\end{figure}
\begin{figure}[!t]
\begin{center}
\vspace{-2.0cm}
\hspace{-0.5cm}
\mbox{\hspace{-0.5cm}
\subfigure[][]{\hspace{-0.2cm}
\includegraphics[height=.25\textheight, angle =0]{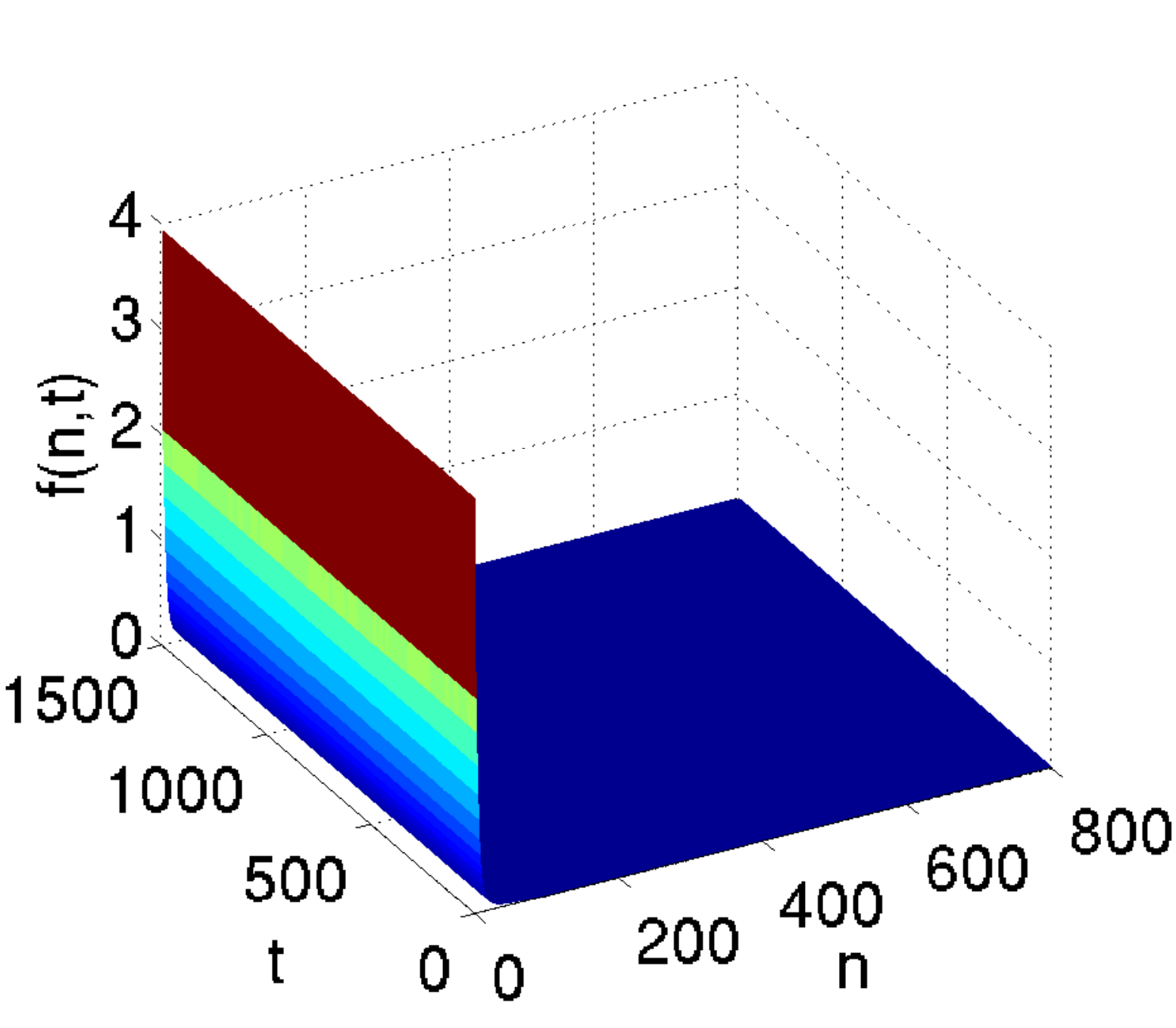}
\label{fig6a}
}
\subfigure[][]{\hspace{-0.2cm}
\includegraphics[height=.24\textheight, angle =0]{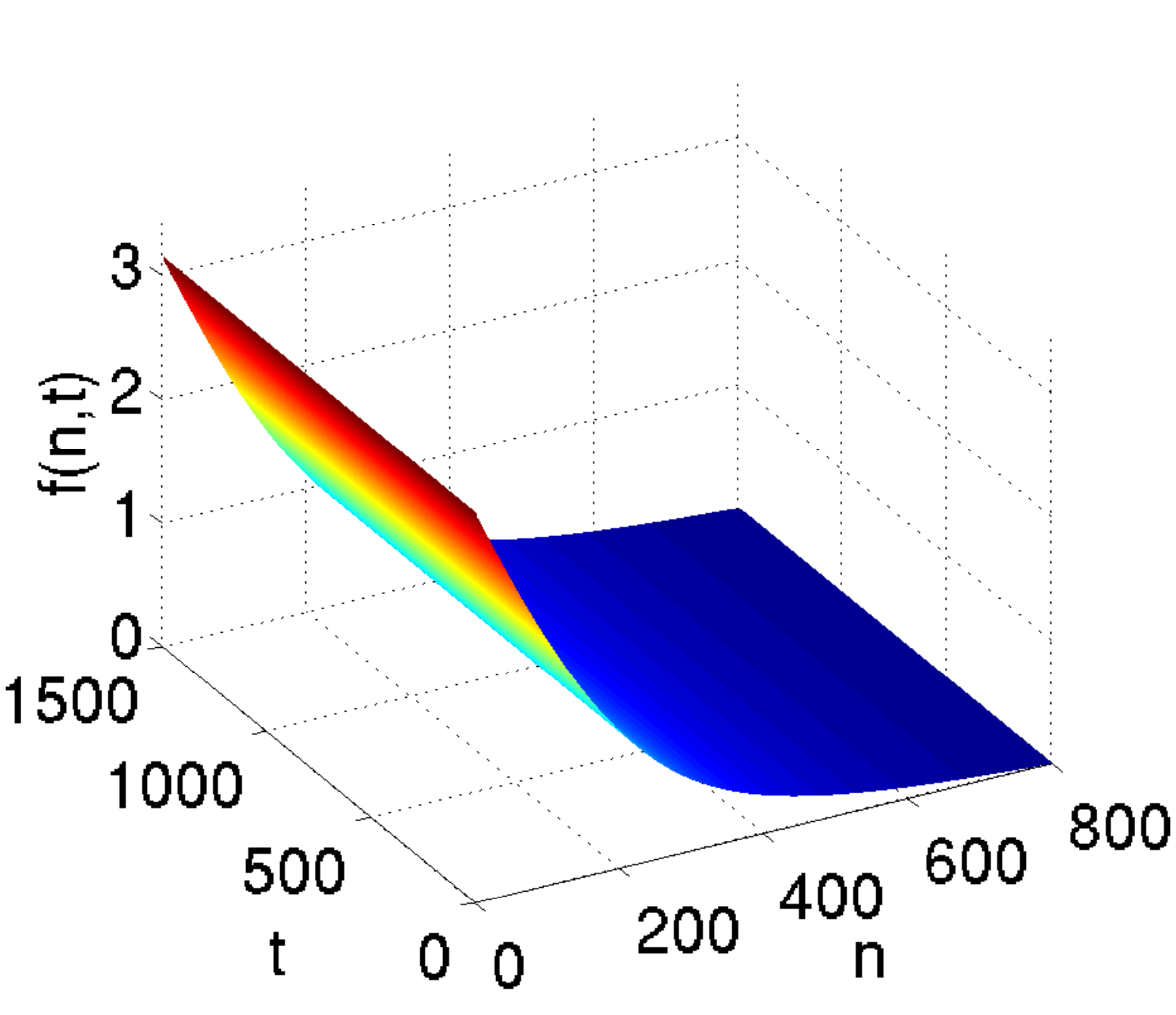}
\label{fig6b}
}
}
\mbox{\hspace{-0.5cm}
\hspace{-0.5cm}
\subfigure[][]{\hspace{-0.2cm}
\includegraphics[height=.24\textheight, angle =0]{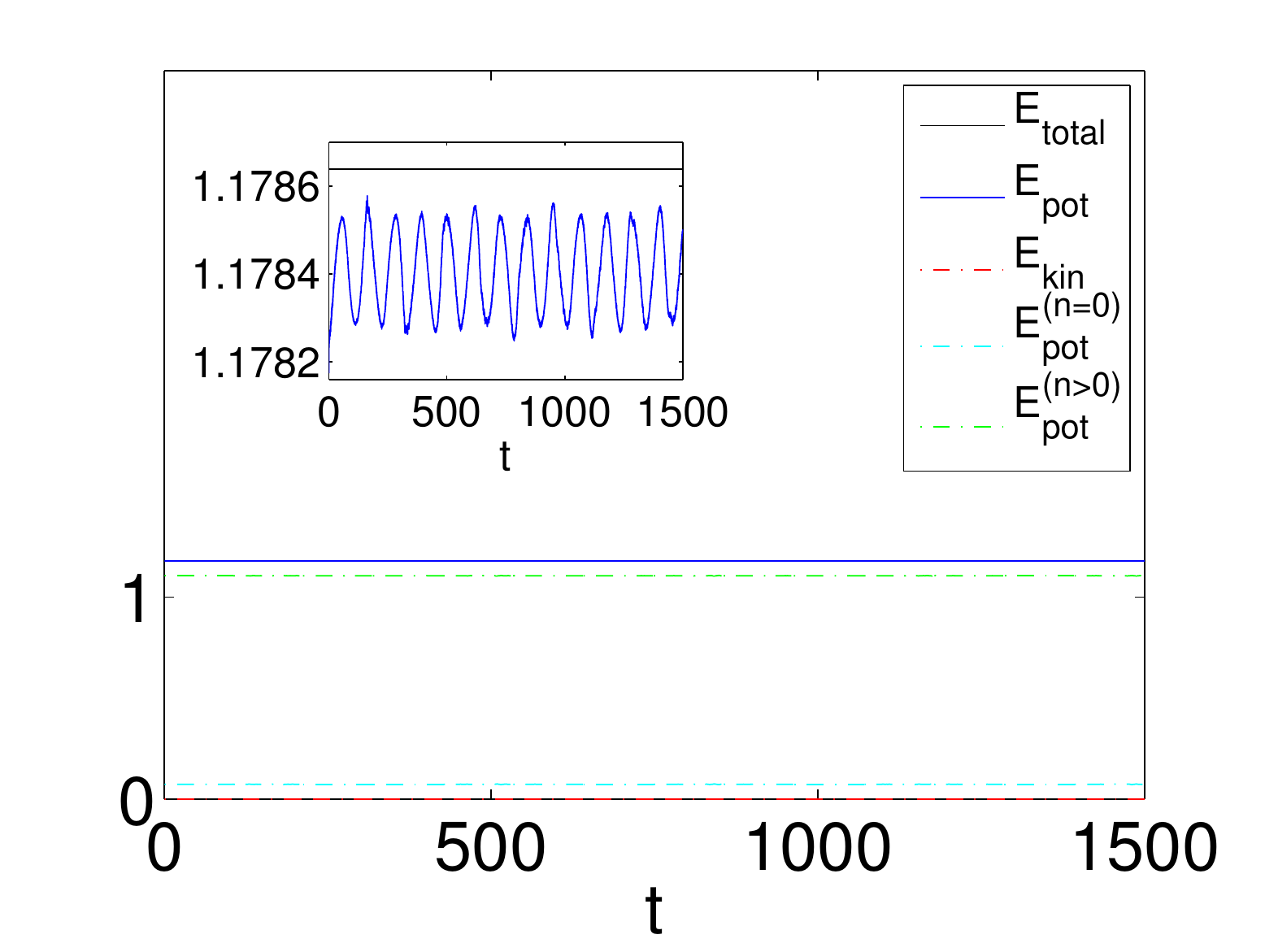}
\label{fig6c}
}
\hspace{-0.5cm}
\subfigure[][]{\hspace{-0.2cm}
\includegraphics[height=.24\textheight, angle =0]{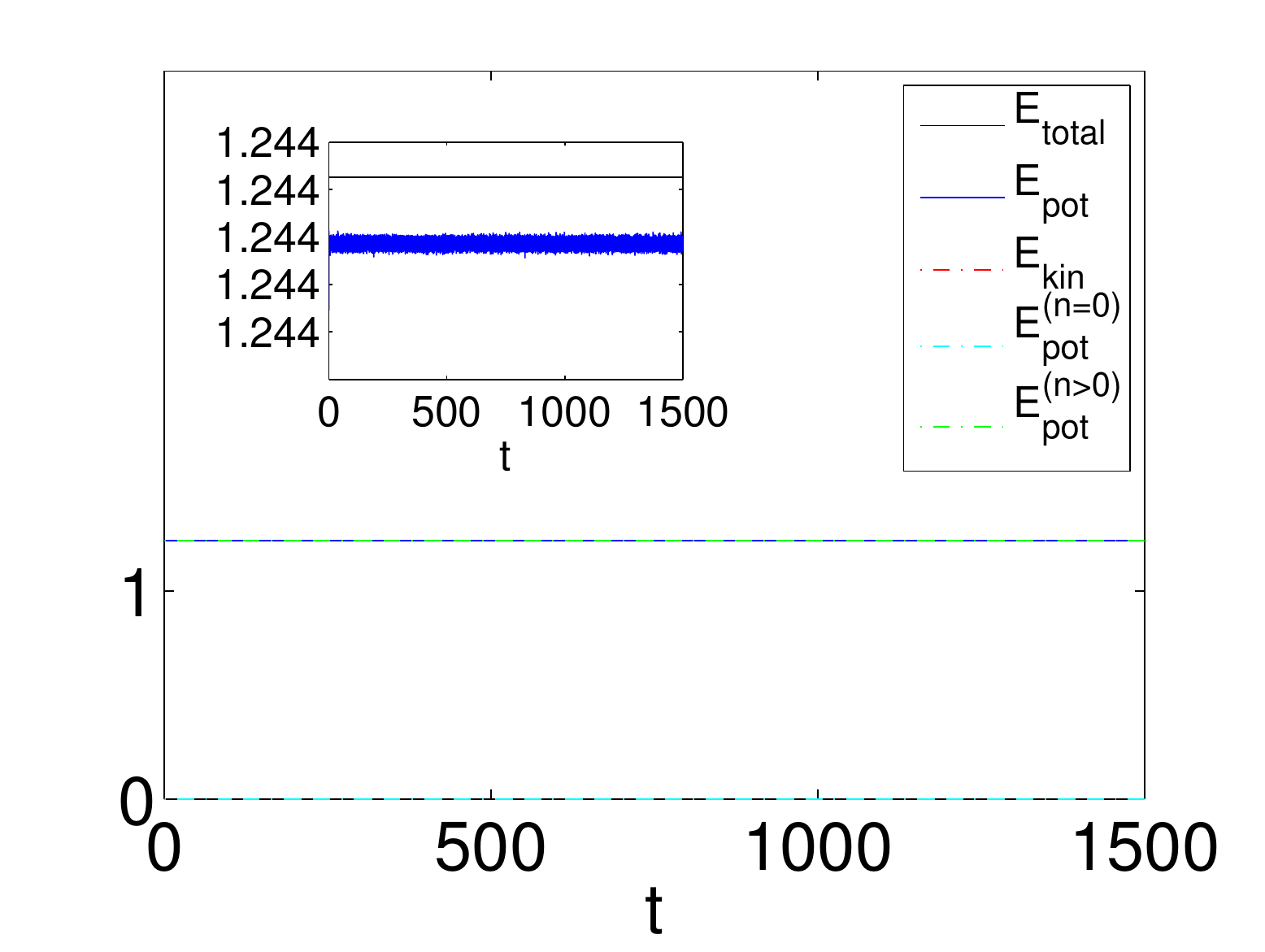}
\label{fig6d}
}
}
\mbox{\hspace{-0.5cm}
\hspace{-0.5cm}
\subfigure[][]{\hspace{-0.2cm}
\includegraphics[height=.245\textheight, angle =0]{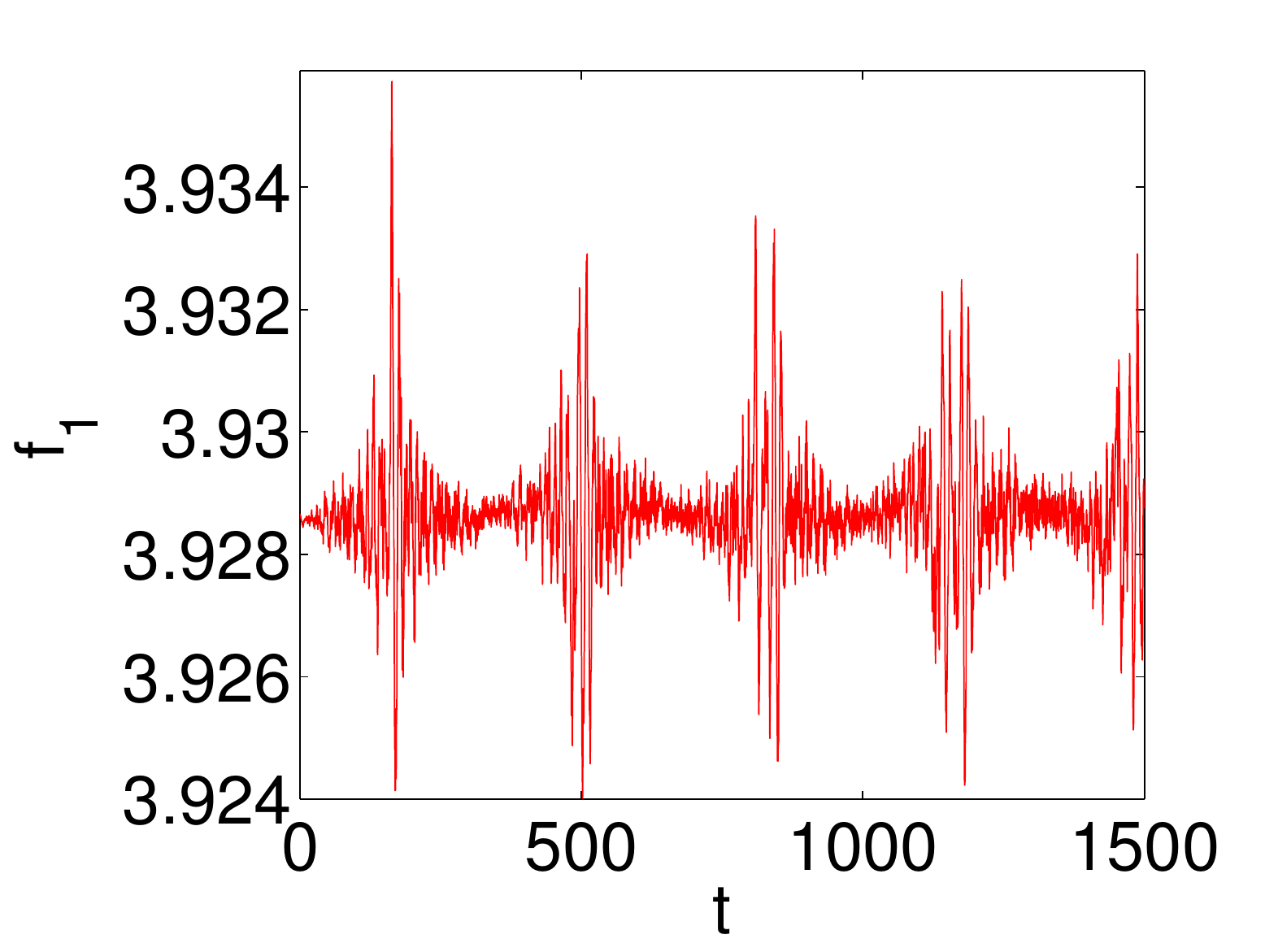}
\label{fig6e}
}
}
\hspace{-0.5cm}
\subfigure[][]{\hspace{-0.2cm}
\includegraphics[height=.245\textheight, angle =0]{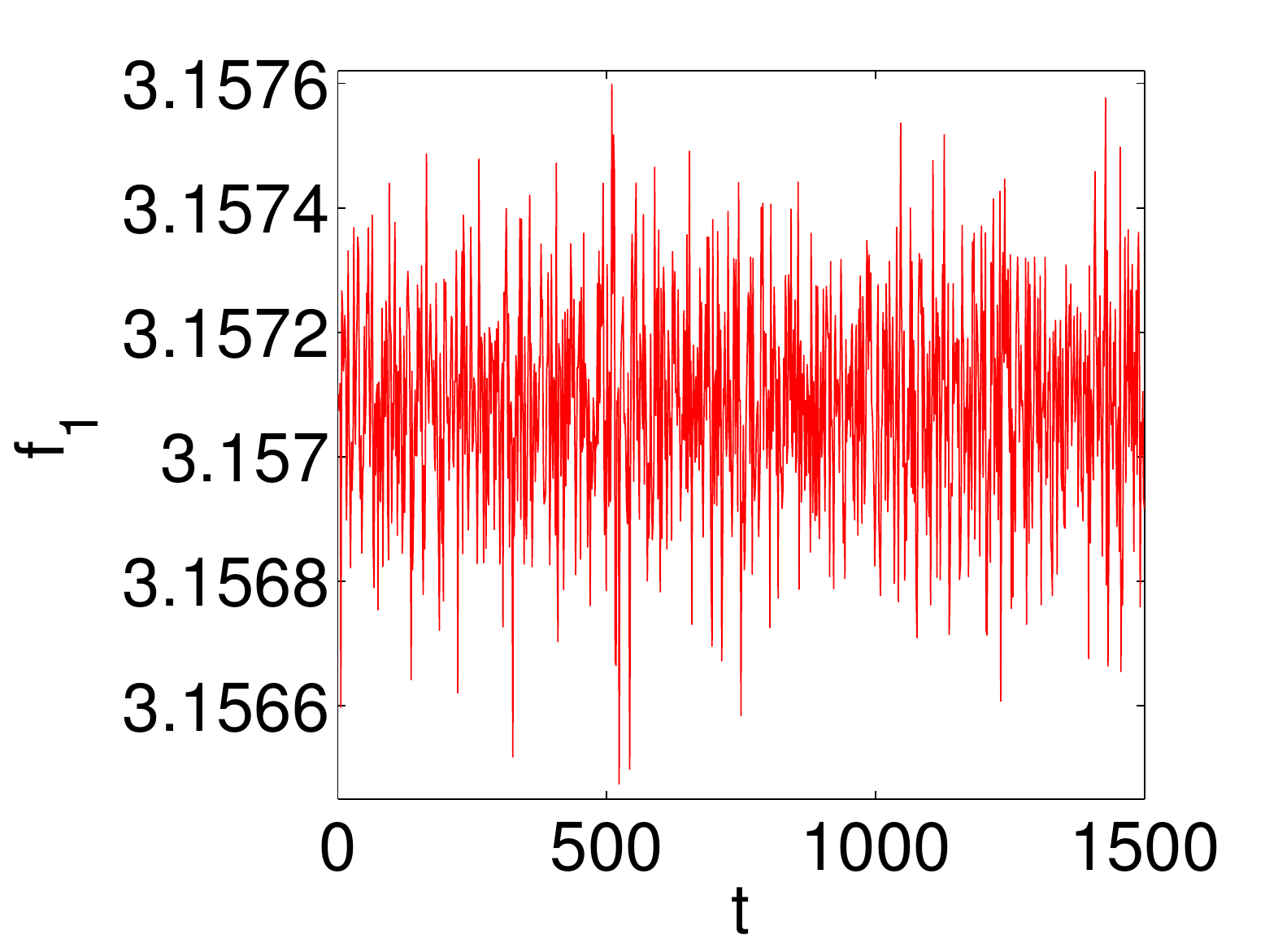}
\label{fig6f}
}
\end{center}
\caption{(Color online) Time evolution of the perturbed solutions of 
Fig.~\ref{fig2}. Left and right panels correspond to values of the 
lattice spacing $h=0.2$ and $h=0.006$, respectively. The top panels 
(a-b) show the space-time evolution of the profile function $f(n,t)$. 
The middle panels (c-d) illustrate the evolution of the kinetic and 
potential energies, and the conservation of the total energy. The bottom
panels (e-f) show the dependence of the profile function $f_1$ versus 
time, again in the presence of a small random initial perturbation.
}
\label{fig6}
\end{figure}

Let us conclude by stating that solutions ``living'' in stable regions of both branches
(and particularly those physically resembling the corresponding skyrmion in the continuum
limit) are natural discrete representations of the corresponding topological solitary 
wave and, in fact, dynamically robust ones such. The latter is illustrated in more detail
in Fig.~\ref{fig5} for solutions of the first branch and in Fig.~\ref{fig6} for those of 
the second one (in connection with the solutions in panels (a) and (b) of Figs.~\ref{fig1}
and \ref{fig2}, respectively). Both figures confirm the dynamical stability of the obtained
discrete solitons by testing the direct dynamical evolution of the system of equations 
(\ref{eom}) in the presence of small, random (uniformly distributed) perturbations imposed
on top of the solitary wave. These perturbations lead solely to oscillations both in the 
kinetic and potential energy of the solutions --while their total energy remains very 
accurately conserved-- and in the ordinate of the first site. To sum up, the solution does
not structurally modify its form for either branch.

\section{Conclusions}

In this brief communication, we have revisited the problem of
three-dimensional skyrmions and the corresponding discretization
of the radial continuum problem in order to pose it on a lattice
setting (and therefore with a natural cutoff scale). We have presented
a self-consistent form of the relevant discretization, amending the
earlier work of~\cite{IK} in that regard, as concerns the energy
of the site which lies immediately next to the origin. 
This discretization has led to the identification of two 
principal branches
of solutions in the form of discrete skyrmions, which both can
be continued down for very small values of $h$. However, only one
is smooth, positive definite and stable,
becoming strongly reminiscent of the continuum limit.
The approach to the limit is further investigated by gradually 
magnifying the radial domain, thus revealing the consistency of
the radial discretization scheme employed. 
Interestingly, a fairly complex bifurcation diagram is revealed
for both branches, including a number of turning points and 
saddle-node bifurcations.
Then, solutions of both 
branches belonging to parametric regions which are found to be linearly
stable (by eigenvalue computations within the realm of linear stability
analysis) have been tested against direct numerical simulations under
the presence of small perturbations.

While alternative discretizations of the radial $SU(2)$ Skyrme 
problem (potentially extended e.g. to the $SU(3)$ one) are of 
interest in their own right, the identification of discretizations
of the axially symmetric problem, or even the fully three-dimensional
one that possess stable discrete skyrmions that naturally
approach their continuum siblings as the lattice spacing tends
to zero, remains a broader problem of interest in its own right.
Studies along this direction are currently in progress and will 
be reported in future publications. 

\vskip 20pt
{\bf Acknowledgements}

The authors acknowledge support from FP7, Marie Curie Actions, People, 
International Research Staff Exchange Scheme (IRSES-605096). E.G.C. is 
indebted to the Institute of Physics, Carl von Ossietzky University 
(Oldenburg) for the kind hospitality provided while part of this work 
was carried out and  acknowledges financial support from the German Research 
Foundation DFG, the DFG Research Training Group 1620 ``{\it Models of Gravity}".
T.A.I. also acknowledges support from The Hellenic Ministry of Education: 
Education and Lifelong Learning Affairs, and European Social Fund: NSRF 2007-2013, 
Aristeia (Excellence) II (TS-3647). P.G.K. also acknowledges support from 
the National Science Foundation under grants CMMI-1000337, DMS-1312856, from
the Binational Science Foundation under grant 2010239 and from the US-AFOSR 
under grant FA9550-12-10332.


\begin{thebibliography}{99}
\bibliographystyle{plain}

\bibitem{remoiss} M. Remoissenet, 
{\it Waves called solitons} (Springer, Berlin, 1999).

\bibitem{nesterenko} V.F. Nesterenko,
{\it Dynamics of Heterogeneous Materials}, 
Springer-Verlag (New York, 2001).

\bibitem{dauxois} T. Dauxois and M. Peyrard,
{\it Physics of Solitons}, Cambridge University Press,
(Cambridge, 2006).

\bibitem{IPA} 
T. Ioannidou, J. Pouget and E. Aifantis, J. Phys. A {\bf 36} 645 (2003); J. Phys. A {\bf 34}  (2001) 4269. 

\bibitem{SW}
J. M. Speight and R. S. Ward, Nonlinearity {\bf 7} (1994) 475.

\bibitem{L}
R. A. Leese, Phys. Rev. D {\bf40} (1989) 2004.

\bibitem{I}
T. Ioannidou,  Nonlinearity {\bf 10} (1997) 1357.

\bibitem{S}
J.M. Speight, Nonlinearity {\bf 10} (1997) 1615; Nonlinearity {\bf12} (1999) 1373. 

\bibitem{Z}
W.Z. Zakrzewski, Nonlinearity {\bf 8} (1995) 517.


\bibitem{IK}
T. Ioannidou and P. Kevrekidis, Phys. Lett. A {\bf372} (2008) 6735.

\bibitem{IKV}
T. Ioannidou, V. B. Kopeliovich and N.D. Vlachos, Nucl. Phys. B
 {\bf 660} (2003) 156.

\bibitem{Ward}
R.S. Ward, Commun. Math. Phys. {\bf 184} (1997) 397.


\bibitem{THR} 
T.H.R Skyrme, Proc. Roy. Soc.  {\bf 260} (1961) 127.


\bibitem{HMS}
C. Houghton, N. Manton and P. Sutcliffe, Nucl. Phys. B \textbf{510} (1998) 587.

\bibitem{COLSYS}
U. Ascher, J. Christiansen and R.D. Russell, Math. Comput. \textbf{33} (1979) 659; %
ACM Trans. Math. Softw. \textbf{7} (1981) 209.

\bibitem{CIM}
E.G. Charalampidis, T.A. Ioannidou and N.S. Manton, J. Math. Phys. \textbf{52} (2011) 033509.


\bibitem{doedel_I} E. Doedel, H.B. Keller and J.P. Kern\'evez, Internat. %
                   J. Bifur. and Chaos \textbf{01} (1991) 493.

                   
\bibitem{AUTO} E. Doedel, AUTO, \textit{indy.cs.concordia.ca/auto/}.                    


\end{thebibliography}
\end{document}